\documentclass[letter,11pt]{article}
\usepackage{jheppub}
\usepackage{tcolorbox}
\usepackage{hyperref}
\usepackage{enumerate}
\usepackage{subcaption}

\tcbuselibrary{theorems}
\allowdisplaybreaks
\subheader{\begin{flushright}
		UTTG-05-2020
\end{flushright}}
\title{Entanglement entropy in cubic gravitational theories}

\author[a]{Elena C\'aceres,}
\author[a]{Rodrigo Castillo V\'asquez,}
\author[b,c]{Alejandro Vilar L\'opez}

\affiliation[a]{Theory Group, Department of Physics, University of Texas, Austin, TX 78712, USA}
\affiliation[b]{Departamento de F\'isica de Part\'iculas, Universidade de Santiago de Compostela, E-15782 Santiago de Compostela, Spain}
\affiliation[c]{Instituto Galego de F\'isica de Altas Enerx\'ias (IGFAE), Universidade de Santiago de Compostela, E-15782 Santiago de Compostela, Spain}
\emailAdd{elenac@utexas.edu}
\emailAdd{rcastillov@utexas.edu}
\emailAdd{alejandrovilar.lopez@usc.es}
\abstract{We derive the holographic entanglement entropy functional for a generic gravitational theory whose action contains terms up to cubic order in the Riemann tensor, and in any dimension. This is the simplest case for which the so-called splitting problem manifests itself, and we explicitly show that the two common splittings present in the literature - minimal and non-minimal - produce different functionals. We apply our results to the particular examples of a boundary disk and a boundary strip in a state dual to 4-dimensional Poincar\'e AdS in Einsteinian Cubic Gravity, obtaining the bulk entanglement surface for both functionals and finding that causal wedge inclusion is respected for both splittings and a wide range of values of the cubic coupling.}

\newtcbtheorem[number within=section]{result}{Result}%
{colback=blue!5,colframe=blue!35!black,fonttitle=\bfseries,before skip=12pt,after skip=12pt}{res}
\newtcbtheorem[number within=section]{question}{Open question}%
{colback=green!5,colframe=green!35!black,fonttitle=\bfseries,before skip=12pt,after skip=12pt}{question}
\newtcbtheorem[number within=section]{warning}{Warning}%
{colback=red!10,colframe=red!50!black,fonttitle=\bfseries,before skip=12pt,after skip=12pt}{warning}

\def \bz {\bar{z}}

\begin{document}

\maketitle
\flushbottom
\newpage

\section{Introduction}
\label{sec:Introduction}

Classical stringy corrections lead to an effective higher-derivative theory of gravity. In such a theory, if the  higher-derivative operators are suppressed by powers of $\ell_P$, we are  guaranteed that the theory is well behaved. When the higher-derivative operators are unsuppressed we have to analyze each theory individually; general statements regarding the well-posedness of the theory are, alas, hard to come by. Even a  fundamental property as causality has to be re-examined. The generic existence of superluminal modes implies that causality and hyperbolicity of the equations of motion are not guaranteed, and the analysis has to be carried out for each specific theory -- or class of theories \cite{Izumi:2014loa,Reall:2014pwa}.

Despite, or maybe because of, all these characteristics, higher-derivative theories are interesting in more than one way. In holography they provide a testing ground where to understand more deeply how holography works in theories whose duals are more generic CFTs (different central charges, non-supersymmetric, etc). Crucial holographic constructs like the  holographic entanglement entropy are modified in the case of higher derivative theories \cite{Hung:2011xb,Dong:entanglement_entropy,Camps:generalized_entropy}, and certain subtleties arise \cite{CampsBoundaryCones}. We will discuss this in detail in the next sections. The generic presence of superluminal modes in higher-derivative theories implies that previous causal constructs based on null rays have to be reexamined \cite{Caceres:2019pok}. To add to the multitude of ways in which higher-derivative  theories differ from Einstein gravity in the holographic context, there is also the question if a given higher-derivative theory is UV complete and, thus, expected to have a sensible field theory dual, or not. Holographically, relationships like causal wedge inclusion \cite{Headrick:2014cta} and entanglement wedge nesting \cite{Akers:2016ugt} are necessary conditions for a background to have a field theory dual and, thus, can  be used to  rule out certain higher-derivative theories from having unitary relativistic QFT duals. In \cite{Caceres:2019pok} the authors showed how causal wedge inclusion can be used to arrive to the same conclusion as \cite{Camanho:2014apa}.

In the vast landscape of higher-derivative gravities, Lovelock theories are among the most studied; they have the advantage that they yield second order equations of motion. Among them, the quadratic theory,  Gauss-Bonnet, has served as a prototype for many phenomena not present in Einstein gravity. From the violation of the $\eta/s$ bound \cite{Brigante_2008,Buchel_2009}, to recent work related to the information paradox \cite{Chen:2020uac}, Gauss-Bonnet theory has taught us important lessons for holography. Cubic theories of gravity have been studied in the general relativity community \cite{Oliva:2010eb,PablosECG}. Some of their holographic properties have been explored \cite{Myers:2010jv,PablosHolography}, but the explicit form of the entanglement entropy functional, a fundamental quantity in holography, was not known.\footnote{There has been previous work on some particular cubic theories, such as quasi-topological gravity \cite{Bhattacharyya:2013gra,Ghodsi:2015gna}. In any case, the general cubic functional was not known, and the ``splitting problem'' - which we discuss in section \ref{sec:HHEinHDG} - was overlooked.} In this paper we advance the understanding of cubic theories in a holographic context by deriving the holographic entanglement functional for a generic cubic gravity theory. This functional can be applied to cubic Lovelock, quasi-topological gravity and Einsteinian cubic gravity theories. Our result is applicable to general cubic gravity theories in any dimension, with the only restriction that the action does not involve derivatives of the curvature tensors. Obtaining the functional presents subtleties absent in quadratic theories. In \cite{MiaoSplitting1,CampsKelly,MiaoSplitting2,CampsBoundaryCones}, the authors showed that, in general, there is an ambiguity in the calculation of the entanglement entropy functional. This ambiguity, known as the ``splitting problem'', is related to the regularization of the action near the conical singularity that appears in the Lewkowycz-Maldacena prescription \cite{LewkowyczMaldacena}. We investigate this issue in detail and present the functional using two different splittings that we refer to as ``minimal'' and ``non-minimal''. The ``non-minimal'' prescription is known to be correct at the perturbative level. At finite coupling, determining the correct splitting is an open question. However, we illustrate our result calculating the HEE surface in a theory that was built to avoid causality problems at the perturbative level, Einsteinian Cubic Gravity (ECG).

The structure of the paper is as follows. In section \ref{sec:HHEinHDG} we review the general framework for calculating the holographic entanglement entropy in higher-derivative theories. We pay particular attention to the \emph{spliting problem} and to the two different proposals that exist in the literature to solve it. Section \ref{sec:FunctionalCubic} contains our main result: we derive the entanglement entropy functional for a general cubic gravity theory. We explore the result obtained using the two different splitting prescriptions. We point out that quadratic theories are insensitive to the differences between them, as also are Lovelock theories, for which the Jacobson-Myers functional is valid \cite{Jacobson:1993xs}. However, for a generic cubic theory, the minimal and non-minimal prescriptions lead to different answers. As an example of our results, in section \ref{sec:StripECG} we work out in detail the entanglement functional for a particular cubic gravity theory, Einstenian Cubic Gravity (ECG) \cite{PablosECG}, and present some numerical results regarding the minimal surface they produce. Finally, in section \ref{sec:conclusions} we summarize our results, its implications, and point out open directions.

\section{Holographic entanglement entropy in higher-derivative gravity}
\label{sec:HHEinHDG}

In a holographic CFT dual to Einstein gravity, the entanglement entropy of a boundary region $A$ is given by the area of an associated codimension-2 surface \cite{RyuTakayanagi1,RyuTakayanagi2}:
\begin{equation} \label{HEEinHDG:RTFormula}
S = \frac{\text{Area}(\gamma_A)}{4 G_N} ~ .
\end{equation}
The surface $\gamma_A$, also known as the  Ryu-Takayangi  or RT surface,  is defined as the bulk codimension-2 surface which has the minimal area among all those homologous to the region $A$ in the boundary (it has to end in $\partial A$ if this is not empty).  If we assume holography holds, the work of Lewkowycz and Maldacena \cite{LewkowyczMaldacena} constitutes a proof  that \eqref{HEEinHDG:RTFormula}  indeed gives the entanglement entropy. We will briefly review their argument here in order to set the stage for our future discussion concerning field theories dual to higher-derivative gravities. The computation starts by considering the usual replica trick in the boundary field theory. The entanglement entropy $S$ can be computed as the limit $n \to 1$ of the R\'enyi entropies:
\begin{equation} \label{HEEinHDG:RenyiEntropies}
S_n (A) = - \frac{1}{n - 1} \log \text{Tr} \left( \rho_A^n \right) = - \frac{1}{n-1} \left( \log \mathcal{Z}_n - n \log \mathcal{Z}_1 \right)  ~,
\end{equation}
where $\rho_A$ is the reduced density matrix of the subsystem associated with region $A$, and $\mathcal{Z}_n$ is the partition function of the field theory in the $n$-fold cover. This is a manifold consisting of $n$ copies of the original one, glued cyclically at the spatial region $A$. $\mathcal{Z}_1$ is thus just the original partition function. We assume always that an analytic continuation to Euclidean signature has been performed. Notice also that R\'enyi entropies are defined for $n \in \mathbb{N}^{\star}$, therefore an analytic continuation in $n$ is also assumed before taking the limit $n \to 1$.

So far, all this discussion has been restricted to the field theory, but if this is holographically dual to a gravitational one, it should be possible to find a bulk solution $B_n$ dual to the $n$-fold cover. Then, $\log \mathcal{Z}_n = - \mathcal{I}[B_n]$, where $\mathcal{I}[B_n]$ is the on-shell gravitational Euclidean action of this dual geometry. Naturally, $\log \mathcal{Z}_1 = - \mathcal{I}[B_1]$, $B_1$ being just the original bulk dual. Now, the $n$-fold cover boundary manifold has a $\mathbb{Z}_n$ symmetry, due to the fact that we can do permutations on the $n$ copies of the original manifold. If we assume that this replica symmetry is respected in the bulk, we can consider the manifold $\hat{B}_n = B_n / \mathbb{Z}_n$, which has to be regular everywhere except in the codimension-2 submanifold $\mathcal{C}_n$ consisting of fixed points of the $\mathbb{Z}_n$. Notice that, in the boundary, $\partial A$ are precisely the fixed points of $\mathbb{Z}_n$. Also, since $B_n$ is a regular bulk solution, the orbifold $\hat{B}_n$ has a conical defect of opening angle $2 \pi / n$ at $\mathcal{C}_n$. Due to the replica symmetry, we can write:
\begin{equation} \label{HEEinHDG:replicaSymmetryAction}
\mathcal{I}[B_n] = n \mathcal{I}[\hat{B}_n] ~ ,
\end{equation}
where in $\mathcal{I}[\hat{B}_n]$ we exclude contributions coming from the conical singularity (this is because in the left-hand side of the previous equation there are no such contributions, the geometry is regular).\footnote{Boundary terms at the asymptotic boundary where the field theory lives should be included as usual, since they must appear also in $\mathcal{I}[B_n]$.} After doing a suitable analytic continuation of this $\hat{B}_n$ to non-integer $n$, we can finally write:
\begin{equation} \label{HEEinHDG:replicaTrickEntanglement}
S = \lim_{n \to 1} \frac{n}{n-1} \left( \mathcal{I}[\hat{B}_n] - \mathcal{I}[B_1] \right) = \left. \partial_n \mathcal{I}[\hat{B}_n] \right|_{n = 1} ~ ,
\end{equation}
where $\mathcal{I}[\hat{B}_1] = \mathcal{I}[B_1]$. Once this expression is obtained, the computation of the entanglement entropy of the region $A$ has been reduced to a problem in classical gravity, which can be solved in two steps:
\begin{enumerate}
\item The geometry with $n = 1$, $\hat{B}_1$, is a regular solution of the equations of motion. In \eqref{HEEinHDG:replicaTrickEntanglement} we seem to be doing a first order variation away from this solution, so we could naively expect that expression to vanish. This is not so because, when varying, we are changing the opening angle at $\mathcal{C}_1 = \lim_{n \to 1} \mathcal{C}_n$, which as mentioned should be excluded from the action integral and that procedure introduces a boundary where conditions are changing if we vary $n$. This localizes the computation of the entanglement entropy in $\mathcal{C}_1$, and in fact in Einstein gravity it is possible to prove from the form of the action (see \cite{LewkowyczMaldacena}) that $S$ is computed as shown in \eqref{HEEinHDG:RTFormula}. $\gamma_A$ should be interpreted at this point as $\mathcal{C}_1$, where we have not proven its minimal property yet.
\item The remaining question is how we determine $\mathcal{C}_1$. Formally, it is defined by looking for $\mathcal{C}_n$ in the analytically continued spacetime $\hat{B}_n$, and then taking the limit $n \to 1$. Adopting adapted coordinates at the conical singularity (see appendix of \cite{LewkowyczMaldacena}), it is possible to show that the equations of motion derived from Einstein gravity for $\hat{B}_n$ impose, in the limit $n \to 1$, the minimal area condition:
\begin{equation} \label{HEEinHDG:minimalAreaEquations}
K^{a} = 0  ~,
\end{equation}
where $K^a$ are the traces of the extrinsic curvatures along the transverse directions to $\mathcal{C}_1$, and $a$ is an index which runs in these two directions (this notation will be clarified in the following section). This shows that $\mathcal{C}_1$ is a minimal area surface, which can then be calculated by minimization of the entanglement entropy functional \eqref{HEEinHDG:RTFormula}. With this condition, $\mathcal{C}_1$ can be characterized as the previously defined surface $\gamma_A$ which is homologous to the boundary region $A$.
\end{enumerate}
%

\subsection{The entropy functional and the splitting problem}
\label{subsec:SplittingProblem}

The previous program can be carried out, with an increasing level of technical difficulty, when the gravitational theory contains higher-derivative corrections to the Einstein-Hilbert action. Following the two steps we have just described, \cite{Dong:entanglement_entropy} and \cite{Camps:generalized_entropy} first obtained the expression for the functional computing entanglement entropy in the presence of higher-derivative terms. Using \eqref{HEEinHDG:replicaTrickEntanglement}, and considering a Lagrangian containing arbitrary contractions of the Riemann tensor (but not its derivatives), one obtains:
\begin{equation} \label{SplittingProblem:general_functional}
S_{EE} = 2 \pi \int_{\mathcal{C}_1} d^{D-2} y \, \sqrt{g} \, \left[ \frac{\partial \mathcal{L}_E}{\partial R_{z \bz z \bz}} + \sum_{A} \left( \frac{\partial^2 \mathcal{L}_E}{\partial R_{z i z j} \partial R_{\bz k \bz l}} \right)_A \frac{8 K_{zij} K_{\bz k l}}{q_A + 1} \right] ~ .
\end{equation}
where $\mathcal{L}_E = \mathcal{L}_E \left( R_{\mu \nu \rho \sigma} \right)$ is the Euclidean version of the Lagrangian. Let us explain the notation in this expression, which employs the conventions of \cite{Dong:entanglement_entropy}:
\begin{itemize}
\item The full manifold has dimension $D$, and we denote generic coordinates in it by $x^{\mu}$. Indices for this $D$-dimensional manifold are $\mu, \nu, \rho, \sigma, \dots$ The surface on which the previous functional is evaluated, $\mathcal{C}_1$, is $(D-2)$-dimensional. Coordinates in it are $y^i$, and indices will be labelled $i, j, k, l, \dots$ We assume an embedding $x^{\mu} = x^{\mu}(y^i)$, so that we can define tangent vectors $(m_i)^{\mu} \equiv \partial_{i} x^{\mu}$, and then take two extra orthonormal vectors $n_a$ to complete the basis, $G_{\mu \nu} (n_a)^\mu (n_b)^\nu = \delta_{ab}$. Indices $a, b, c, d, \dots$ will be used for these two directions.
\item The functional \eqref{SplittingProblem:general_functional} is defined using a particular set of adapted coordinates for $\mathcal{C}_1$ (see \cite{Dong:entanglement_entropy}), where tangent coordinates are $x^i(y) = y^i$, and we introduce two extra complex coordinates $z, \bz$ such that the metric factorizes:
\begin{equation} \label{SplittingProblem:metric_surface}
\left. G_{z \bz} \right|_{\mathcal{C}_1} = \frac{1}{2} , \qquad \left. G_{i j} \right|_{\mathcal{C}_1} = g_{i j} , \qquad \left. G^{z \bz} \right|_{\mathcal{C}_1} = 2 , \qquad \left. G^{i j} \right|_{\mathcal{C}_1} = g^{i j} ~ ,
\end{equation}
with the remaining components vanishing at $\mathcal{C}_1$.
\item $K^a{}_{ij}$ is the extrinsic curvature of $\mathcal{C}_1$ along the direction $n^a$:
\begin{equation} \label{SplittingProblem:definition_K}
K^a{}_{ij} \equiv (m_i)^\mu (m_j)^\nu \nabla_{\mu} (n^a)_{\nu} = - (n^a)_\mu \left[ \partial_i \partial_j x^{\mu} + \Gamma^\mu_{ \nu \rho} \partial_i x^{\nu} \partial_j x^{\rho} \right] ~ ,
\end{equation}
where $(n^a)_\mu = \delta^{ab} G_{\mu \nu} (n_b)^{\nu}$. This appears in \eqref{SplittingProblem:general_functional} as a spacetime tensor defined in the usual way: $K^{\mu}{}_{\nu \rho} \equiv K^a{}_{ij} (n_a)^{\mu} (m^i)_{\nu} (m^j)_{\rho}$. Notice that, in the adapted coordinates, the orthonormal vectors are going to be taken as:
\begin{equation} \label{SplittingProblem:orthonormal_vectors}
n_1 = \sqrt{\frac{z}{\bz}} \partial_z + \sqrt{\frac{\bz}{z}} \partial_{\bz}  ~, \quad n_2 = i \left( \sqrt{\frac{z}{\bz}} \partial_z - \sqrt{\frac{\bz}{z}} \partial_{\bz} \right) ~ .
\end{equation}
\end{itemize}

There remains to explain the sum over $A$ in the last term of \eqref{SplittingProblem:general_functional}, usually called the \emph{anomaly} term. Its origin is subtle, coming from potentially logarithmically divergent terms in the action at $\mathcal{C}_1$ when taking the limit $n \to 1$. For those terms, a careful analysis of the limit has to be performed, and it becomes essential to understand the analytic continuation of the geometry and the regularization of the conical singularity at $\mathcal{C}_1$.\footnote{This is because the computation of \eqref{HEEinHDG:replicaTrickEntanglement} can be localized at $\mathcal{C}_1$, as already mentioned. See \cite{Dong:entanglement_entropy} for more details.} This regularization has to be done guaranteeing that the equations of motion of the theory are satisfied. This was at first overlooked in \cite{Dong:entanglement_entropy}, where a \emph{minimal} regularization was employed. It is nevertheless useful to quote the result obtained, since it will allow us to see more clearly the differences introduced when other regularizations are used. For the computation of the entropy functional, all the study of the behaviour of the action around $\mathcal{C}_1$ boils down to the following algorithmic procedure. In a general theory, the second derivative of the Lagrangian in the last term of \eqref{SplittingProblem:general_functional} will be a polynomial in curvature tensors. In this polynomial we expand every curvature tensor using the following expressions:
\begin{align} \label{SplittingProblem:expansions_Riemann}
\nonumber R_{\alpha \beta ij} & = \tilde{R}_{\alpha \beta ij} + g^{kl} \left[ K_{\alpha jk} K_{\beta il} - K_{\alpha ik} K_{\beta jl} \right] ~, \\
R_{\alpha i \beta j} & = \tilde{R}_{\alpha i \beta j} + g^{kl} K_{\alpha jk} K_{\beta il} - Q_{\alpha \beta ij}  ~, \\
\nonumber R_{ikjl} & = r_{ikjl} + G^{\alpha \beta} \left[ K_{\alpha il} K_{\beta jk} - K_{\alpha ij} K_{\beta kl} \right] ~,
\end{align}
where indices $\alpha, \beta, \gamma, \dots$ denote values $z$ or $\bz$, $r_{ikjl}$ is the lower-dimensional Riemann tensor, and $\tilde{R}_{\alpha \beta ij}$, $\tilde{R}_{\alpha i \beta j}$ and $Q_{\alpha \beta i j}$ are defined in \cite{Dong:entanglement_entropy} but their particular form will not be needed here. Once the expansion is done, we label each of the individual terms with $A$. Considering any of them, we associate a value $q_A$ to it equal to the number of factors of $Q_{z z i j}$ and $Q_{\bz \bz i j}$ plus one half the number of factors of $K_{\alpha ij}$, $R_{\alpha \beta \gamma i}$ and $R_{\alpha ijk}$. Once this is done, we divide by $q_A + 1$, as indicated in \eqref{SplittingProblem:general_functional}. This expansion is what we denote by the sum over $A$, and when it is completed one can rewrite, if desired, everything again in terms of the original curvature tensors using \eqref{SplittingProblem:expansions_Riemann}.

As mentioned, this prescription (which we will call \emph{minimal} prescription) does not take into account the fact that the regularized metric at the conical singularity has to satisfy the equations of motion when computing \eqref{HEEinHDG:replicaTrickEntanglement}. The existence of several ways to regularize the metric (with relevant consequences for the entropy functional) has been called the \emph{splitting problem} in the literature, and discussions around this issue can be found in \cite{MiaoSplitting1,CampsKelly,MiaoSplitting2}. 

For a general higher-derivative theory, it can be quite complicated to impose the on-shell condition for the regularization, but one can make a first approach to the problem by studying the constraint imposed by Einstein gravity. This produces a functional which is at least perturbatively correct, since the leading order area term is independent of the splitting chosen. This is done in great detail in \cite{MiaoSplitting2} (and in \cite{CampsBoundaryCones}, although the final result is expressed assuming the surface has $K^a = 0$, so that it has extremal area), and it is shown that the final effect for the entropy functional is a different prescription for the $A$ expansion (we will call this \emph{non-minimal} prescription). After expanding all the curvature tensors according to \eqref{SplittingProblem:expansions_Riemann}, we have to rewrite $R_{z \bz z \bz}$ and $Q_{z \bz i j}$ in terms of two new objects: 
\begin{align} \label{SplittingProblem:primed_tensors}
R'_{z \bz z \bz} & = R_{z \bz z \bz} + \frac{1}{2} \left( K_{z ij} K_{\bz}{}^{ij} - K_{z} K_{\bz} \right) ~ , \\
Q'_{z \bz i j} & = Q_{z \bz i j} - K_{z i}{}^k K_{\bz j k} - K_{z j}{}^k K_{\bz i k} + \frac{1}{2} K_{z} K_{\bz i j} + \frac{1}{2} K_{\bz} K_{z i j} ~ ,
\end{align}
and associate $q_{A} = 1/2$ to $K_{\alpha i j}$, $R_{\alpha \beta \gamma i}$, and $R_{\alpha i j k}$; and $q_{A} = 1$ to $Q_{z z i j}$ and $Q_{\bz \bz i j}$. Then we must divide by $q_{A} + 1$ as indicated in the general expression \eqref{SplittingProblem:general_functional}, and finally, if desired, undo the expansions, writing everything in terms of curvature tensors again.

We have presented two ways to define the functional computing entanglement entropy in the presence of higher-derivative corrections to the gravitational action. Let us emphasize that for quadratic theories the functionals obtained using the \emph{minimal} and \emph{non-minimal} prescriptions are the same. As shown in \cite{Dong:entanglement_entropy}, this is also the case for Lovelock theories, where the Jacobson-Myers functional is known to be correct. This is a special property of Lovelock theories, for which the entanglement entropy functional depends only on the intrinsic geometry of the surface, $r_{i j k l}$. However, we will show that for general cubic gravities the functionals obtained are different.

Until now, we have only completed the first step of the general strategy outlined in the previous section, \emph{i.e.}, we have shown how to obtain the entanglement functional starting from the action of the theory. We still have to find the position of the surface $\mathcal{C}_1$, in other words,  where to evaluate the functional. In principle,  the equations of motion of the theory should determine the location of the surface, in much the same way they determine it to be a minimal area surface in General Relativity. The idea is to explicitly evaluate the equations of motion in the conically singular metric for generic $n$ to linear order in $n-1$. One does not expect divergences in the stress-energy tensor, but these generically appear in the metric part of the equations of motion, so cancelling them imposes conditions which determine the position of the surface in the limit $n \to 1$ (in particular, GR equations of motion impose the well-known condition $K^{a} = 0$). Further details can be found in \cite{LewkowyczMaldacena,CampsKelly}. In practice, this procedure has the drawback of requiring to deal with the equations of motion of higher-derivative theories, which can be extremely complicated. Fortunately, in \cite{DongExtremality} it is shown that the same procedure one employs in Einstein gravity is also valid in general: after computing the correct functional, one can minimize it to obtain the surface $\mathcal{C}_1$ in which it is going to be evaluated.\footnote{Notice that the equations of motion are still necessary in principle, because one has to determine the correct splitting. It is only possible to state the correctness of the non-minimal prescription at the perturbative level.}\textsuperscript{,}\footnote{Previous work on the question of whether minimizing the functional is equivalent to the Lewkowycz-Maldacena prescription found some issues for certain theories \cite{Bhattacharyya2014}. This may be due to the fact that the splitting problem is being ignored, but it would be interesting to check it explicitly.}

\section{Entanglement entropy functional in cubic gravity}
\label{sec:FunctionalCubic}

This section contains our main result, we will derive the entanglement functional for a \emph{generic cubic gravity theory} that does not involve explicit derivatives of the curvature tensor. We will use the method diuscussed in section \ref{sec:HHEinHDG} to compute the different contributions to the holographic entanglement entropy functional in cubic gravities.

As a warm-up exercise, let us revisit the known result for quadratic theories. As explained in section \ref{sec:HHEinHDG}, there are different prescriptions to calculate the entanglement entropy functional in higher-derivative theories. In quadratic gravity theories,
\begin{equation} \label{FunctionalCubic:QuadraticL}
\mathcal{L}_E = \lambda_1 R^2 + \lambda_2 R^{\mu \nu} R_{\mu \nu} + \lambda_3 R_{\mu \nu \rho \sigma} R^{\mu \nu \rho \sigma} ~ , 
\end{equation}
any prescription leads to the same result for the holographic entropy functional \cite{FursaevPatrushevSolodukhin}. The reason for this is that for quadratic theories the expansion in $A$ is trivial: after two derivatives of the Lagrangian we obtain something which does not contain curvature tensors, so essentially $q_{A} = 0$ always. This guarantees that, in the case of quadratic gravities,  the result obtained  using either the \emph{minimal} or \emph{non-minimal} splitting is the same,
\begin{align} \label{FunctionalCubic:QuadraticS}
S^{quad}_{EE} = - 4 \pi \int d^d y \sqrt{g} \left[ 2 \lambda_1 R + \lambda_2 \left( R^{a}{}_a - \frac{1}{2}K_a K^a \right) + 2 \lambda_3 \left( R^{a b}{}_{a b}  -  K_{a i j} K^{a i j} \right) \right] ~ ,
\end{align}
where $K_a \equiv K_{aij} g^{ij}$. This is not the case for cubic gravities. Consider the following generic cubic Lagrangian:\footnote{We do not consider terms with explicit derivatives of the curvature tensors, such as $\nabla_{\mu} R \nabla^{\mu} R$. These could in principle appear also at cubic order, but they complicate considerably the calculations, and many of the known cubic gravity theories like Lovelock \cite{Lovelock:1971yv}, quasi-topological gravity \cite{Oliva:2010eb,Myers:2010jv}, and ECG \cite{PablosECG} do not include them. How to deal with these terms can be found in \cite{MiaoSplitting1}.}
\begin{equation} \label{FunctionalCubic:Lagrangian_cubic}
\begin{array}{cl}
\mathcal{L}_E = & \mu_8 R^3 + \mu_7 R_{\mu \nu} R^{\mu \nu} R + \mu_6 R_{\mu}{}^{\nu} R_{\nu}{}^{\rho} R_{\rho}{}^{\mu} + \mu_5 R^{\mu \rho} R^{\nu \sigma} R_{\mu \nu \rho \sigma}+ \mu_4 R_{\mu \nu \rho \sigma} R^{\mu \nu \rho \sigma} R \\ [0.6em]
 & + \mu_3 R^{\mu \nu \rho}{}_{\sigma} R_{\mu \nu \rho \tau} R^{\sigma \tau} + \mu_2 R^{\mu \nu}{}_{\rho \sigma} R^{\rho \sigma}{}_{\lambda \tau} R^{\lambda \tau}{}_{\mu \nu} + \mu_1 R_{\mu}{}^{\rho}{}_{\nu}{}^{\sigma} R_{\rho}{}^{\lambda}{}_{\sigma}{}^{\tau} R_{\lambda}{}^{\mu}{}_{\tau}{}^{\nu} ~ . \quad 
\end{array}
\end{equation} 
Note that the second derivative of  the Lagrangian \eqref{FunctionalCubic:Lagrangian_cubic} is linear in the curvature tensor. Therefore, unlike in quadratic gravity, the two different splittings discussed in section \ref{sec:HHEinHDG} lead to different entanglement functionals. The details of these highly technical calculations are presented  in  Appendix \ref{app:details_func}. The final result for the entropy functional obtained following the \emph{minimal} splitting prescription is:
\begin{equation} \label{FunctionalCubic:functional_XD}
S^{min}_{EE} = 2 \pi \int d^{D-2} y \sqrt{g} \left( S_{R^2} + S_{K^2 R} + S^{min}_{K^4} \right) ~ ,
\end{equation}
where
\begin{flalign} \label{FunctionalCubic:Wald_term}
S_{R^2} = & - 6 \mu_8 R^2 - 2 \mu_7 \left( R_{\mu \nu} R^{\mu \nu} + R_a{}^a R \right) - 3 \mu_6 R_{a \mu} R^{a \mu} - \mu_5 \left( 2 R_{\mu \nu} R_{a}{}^{\mu a \nu} - R_{ab} R^{ab} + R_a{}^a R_b{}^b \right) \nonumber & \\
&  - 2 \mu_4 \left( R_{\mu \nu \rho \sigma} R^{\mu \nu \rho \sigma} + 2 R R_{ab}{}^{ab} \right) - \mu_3 \left( R_{a \mu \nu \rho} R^{a \mu \nu \rho} + 4 R_{a \mu} R_b{}^{a b \mu} \right) \nonumber & \\
& - 6 \mu_2 R_{a b \mu \nu} R^{a b \mu \nu} + 3 \mu_1 \left( R_{a \mu b \nu} R^{a \nu b \mu} - R_{a \mu}{}^{a}{}_{\nu} R_{b}{}^{\mu b \nu} \right) ~ , &
\end{flalign}
\begin{flalign} \label{FunctionalCubic:K2R_term}
S_{K^2 R} = & + \mu_7 K_a K^a R + \frac{3}{2} \mu_6 K_a K^a R_b{}^b + 2 \mu_5 K_a K^{a}{}_{ij} R^{ij} - \frac{1}{2} \mu_5 K_a K^a R_{bc}{}^{bc} + 4 \mu_4 K_{a i j} K^{a i j} R \nonumber & \\
& + 2 \mu_3 K_{a i k} K^a{}_{j}{}^{k} R^{ij} + \mu_3 K_{aij} K^{aij} R_b{}^b + 2 \mu_3 K_a K^{a}{}_{ij} R_{b}{}^{ibj} + 12 \mu_2 K_{a i k} K^a{}_{j}{}^k R_{b}{}^{i b j} \nonumber & \\
& + 3 \mu_1 K_{a i j} K^{a}{}_{kl} R^{ikjl} - \frac{3}{2} \mu_1 K_{a i j} K^{a i j} R_{bc}{}^{bc} + 6 \left( 2 \mu_2 + \mu_1 \right) K_{a i k} K_{b j}{}^{k} R^{a b i j} ~ , &
\end{flalign}
\begin{flalign} \label{FunctionalCubic:K4_term_XD}
S^{min}_{K^4} = & + \frac{1}{2} \mu_7 K_a K^a K_b K^b + \mu_5 K_a K^{a}{}_{ij} \left( K_{b} K^{bij} - 2 K_{b}{}^{i}{}_{k} K^{bjk} \right) \nonumber & \\
& - \frac{1}{4} \left( 6 \mu_7 + 3 \mu_6 - 8 \mu_4 \right) K_a K^a K_{bij} K^{bij} - \frac{1}{2} \left( 12 \mu_4 + \mu_3 - 3 \mu_1 \right) K_{aij} K^{aij} K_{bkl} K^{bkl} \nonumber & \\
& - \frac{3}{2} \left( 4 \mu_2 + 3 \mu_1 \right) K_{a i}{}^j K_{bj}{}^k K^a{}_{k}{}^l K^b{}_l{}^i + \left( - 2 \mu_3 + 3 \mu_1 \right) K_{a i}{}^j K^a{}_j{}^k K_{b k}{}^l K^b{}_l{}^i ~ . &
\end{flalign}
Using the \emph{non-minimal} prescription we obtain:
\begin{equation} \label{FunctionalCubic:functional_JC}
S^{non-min}_{EE} = 2 \pi \int d^{D-2} y \sqrt{g} \left( S_{R^2} + S_{K^2 R} + S^{non-min}_{K^4} \right) ~ ,
\end{equation}
where the first two terms are given by \eqref{FunctionalCubic:Wald_term} and \eqref{FunctionalCubic:K2R_term}, and the last one is:
\begin{flalign} \label{FunctionalCubic:K4_term_JC}
S^{non-min}_{K^4} = & + \frac{1}{4} \left( \mu_5 - 3 \mu_1 \right) K_a K^a K_{b i j} K^{b i j} - \frac{1}{4} \mu_5 K_a K^a K_b K^b + \left( \mu_3 - 6 \mu_2 \right) K_a K^{a}{}_{ij} K_{b}{}^{i}{}_k K^{b j k}  \nonumber & \\
&  - \mu_3 K_a K^a{}_{i j} K_b K^{b i j} + \frac{3}{4} \mu_1 K_{aij} K^{aij} K_{bkl} K^{bkl} + \frac{3}{2} \mu_1 K_{aij} K_{bkl} K^{bij} K^{akl} \nonumber & \\
& - \frac{3}{2} \left( 4 \mu_2 + 3 \mu_1 \right) K_{a i}{}^j K_{bj}{}^k K^a{}_{k}{}^l K^b{}_l{}^i + 3 \left( 4 \mu_2 + \mu_1 \right) K_{a i}{}^j K^a{}_j{}^k K_{b k}{}^l K^b{}_l{}^i ~ . &
\end{flalign}

Equations \eqref{FunctionalCubic:functional_XD} and \eqref{FunctionalCubic:functional_JC} are two of the main results of this paper. Let us pause and take stock of these results. First, it is natural to inquire whether we can identify where the difference in the two functionals comes from. As the separation of the functional in three different parts makes manifest, we observe that terms which are proportional to the square of background curvature tensors are equal in both prescriptions and given by \eqref{FunctionalCubic:Wald_term}. This is because they come from the first term in the general functional \eqref{SplittingProblem:general_functional} (the \emph{Wald} term), which is independent of the splitting. The same thing happens for terms linear in background curvature tensors \eqref{FunctionalCubic:K2R_term}: although these come from the $A$ expansion, they are such that after the rewriting and counting procedures the result is the same for both prescriptions. At the end, differences arise only in $K^4$ terms. Second, as previously mentioned, determining the correct splitting for a generic cubic theory with finite coupling is an open question. It could be a splitting as yet unknown, different from the previous ones. However, after having obtained the entanglement entropy functional using the minimal and non-minimal prescriptions one can turn the question around and ask: if we use these functionals for a cubic theory with a finite coupling, will either of them produce unwanted, unphysical, behaviour?  In the next section we set out to do just that, we investigate a fundamental property  known as  \emph{causal wedge inclusion}. This property states that, in a spacetime that has a CFT dual, the causal wedge is completely contained in the entanglement wedge, $\mathcal{C}(A) \subseteq \mathcal{E}(A)$. Casual wedge inclusion can be used as a criterion to constrain the space of theories with CFT duals. Studying the causal structure of a generic  higher-derivative theory is usually a thorny issue because in these theories gravity can travel slower or faster than light. The causal structure is determined by characteristic hypersurfaces that are generically non-null. A thorough study of this issue was carried out for Gauss-Bonnet theory in \cite{Izumi:2014loa,Reall:2014pwa}, but the understanding of causality and hyperbolicity properties in cubic theories is an open problem. Therefore, when investigating causal wedge inclusion we will restrict ourselves to a case where we are guaranteed that the causal structure is given by null rays.

\section{A concrete  example: Einsteinian Cubic Gravity (ECG)}
\label{sec:StripECG}

In the previous section we derived the entanglement entropy functional for a generic cubic gravity theory in $D$ dimensions with the only restriction that the action does not involve derivatives of the curvature tensors. Many such theories exist in the literature \cite{Oliva:2010eb,PablosECG}. In this section we will focus on one of these theories to carry out a concrete calculation in full detail. We will compute the entanglement entropy functional of both a disk and a strip in the state dual to vacuum AdS in 4-dimensional Einsteinian Cubic Gravity (ECG), and  explore properties of the entanglement surface.

\subsection{Preliminaries}
\label{sec:ReviewECG}

Before proceeding with the calculation of the entanglement functional in ECG, let us review some aspects of this theory. We refer the reader to the original works \cite{BlackHolesHennigarMann, BlackHolesPablos, PablosHolography} for further details.

\vspace{0.4em}
\noindent\underline{\emph{Einsteinian Cubic Gravity}}
\vspace{0.2em}

\noindent Einsteinian Cubic Gravity (ECG) \cite{PablosECG} is the unique theory containing corrections up to cubic order in the Riemann tensor such that:
\begin{enumerate}
	\itemsep-0.3em 
	\item It has the same spectrum than Einstein gravity when linearized on a maximally symmetric background, that is, it propagates a single transverse and massless graviton.
	\item The coefficients appearing in the Lagrangian are dimension-independent. 
	\item The cubic correction is neither trivial nor topological in four dimensions. 
\end{enumerate}
The action of ECG is 
\begin{equation} \label{Review:Action}
\mathcal{I} = - \frac{1}{16 \pi G_N} \int \, \mathrm{d}^4 x \sqrt{G} \, \left[ R + \frac{6}{L^2} - \frac{\mu L^4}{8} \mathcal{P} \right] ~,
\end{equation}
where
\begin{equation} \label{Review:PLagrangian}
\mathcal{P} = 12 R_{\mu}{}^{\rho}{}_{\nu}{}^{\sigma} R_{\rho}{}^{\lambda}{}_{\sigma}{}^{\tau} R_{\lambda}{}^{\mu}{}_{\tau}{}^{\nu} + R^{\mu \nu}{}_{\rho \sigma} R^{\rho \sigma}{}_{\lambda \tau} R^{\lambda \tau}{}_{\mu \nu} - 12 R^{\mu \rho} R^{\nu \sigma} R_{\mu \nu \rho \sigma} + 8 R_{\mu}{}^{\nu} R_{\nu}{}^{\rho} R_{\rho}{}^{\mu} ~ .
\end{equation}
A similar construction exists for  higher orders in the derivative expansion \cite{AspectsHOG}. Black holes solutions  in this theory exist in the literature and some of their properties have been studied in  \cite{BlackHolesHennigarMann,BlackHolesPablos}.

If we demand stable $\text{AdS}_4$ vacua, the coupling $\mu$ is constrained to a  range of values determined by the equation
\begin{equation} \label{Review:Finfinity}
 1-f_{\infty}+\mu f_{\infty}^3 = 0 ~,
\end{equation}
where $f_\infty$ relates the curvature scale of the AdS background $L_{\star}$ and the action length scale $L$, $L_{\star}^{-2} = f_{\infty} L^{-2}$. To have $\text{AdS}_4$ vacua, the roots of \eqref{Review:Finfinity} should be  $f_{\infty} > 0$, and thus $\mu \leq \frac{4}{27}$. The root $f_{\infty}$ which produces a stable AdS vacuum is
\begin{equation}\label{ReviewECG:Finfinity}
f_{\infty} = \frac{2}{\sqrt{3 \mu}} \sin \left[\frac{1}{3} \arcsin \left(\sqrt{\frac{27 \mu}{4}} \right)\right].
\end{equation}
For $\mu< 0$, there is a single stable (i.e., positive effective Newton constant) vacuum, but it is not possible to have black hole solutions. For  $0\leq \mu<\frac{4}{27}$, there is one stable vacuum connected to Einstein gravity in the limit $\mu \to 0$, and it is possible to have black hole solutions. This root satisfies $f_{\infty}^2 < \frac{1}{3\mu}$.\footnote{A second, positive root of the characteristic equation with $f_{\infty}^2 > \frac{1}{3\mu}$ gives rise to an unstable vacuum, in which the effective Newton constant becomes negative. The case $\mu = \frac{4}{27}$, $f_{\infty}^2 = \frac{1}{3\mu} = \frac{9}{4}$ is a critical limit of the theory, in which the roots merge and the effective Newton constant blows up. We will not consider it here.} Finally, holographic studies show that positivity of energy fluxes at null infinity in the CFT imposes a more stringent constraint in the coupling: $-0.00322 \leq \mu \leq 0.00312$. In \cite{PablosHolography}, the authors argued that the holographic dual of ECG is a non-supersymmetric CFT in three dimensions,\footnote{One of the parameters characterizing the three-point function of the stress tensor, $t_4$, is shown to be different from zero, contrary to what happens in a supersymmetric CFT.} and obtained some entries of the holographic dictionary.

\vspace{0.4em}
\noindent \underline{\emph{Causal wedge inclusion}}
\vspace{0.2em}

\noindent Let us denote $D[\mathcal{A}]$ the causal diamond of a boundary  region $\mathcal{A}$. The causal wedge, $C(\mathcal{A}) $, is the bulk region causally connected to $D[\mathcal{A}]$. That is,  $C(\mathcal{A})$  is the region of  spacetime where there are causal curves that start and end in $D[\mathcal{A}]$. On the other hand, the entanglement wedge, $\mathcal{E}(\mathcal{A})$, defines the region of the bulk that consists of all points spacelike related to the RT (or HRT) surface \cite{Czech:2012bh,Headrick:2014cta}. Subregion duality and AdS/CFT imply a constraint between these two holographic constructs: the causal wedge is completely contained in the entanglement wedge, $C(\mathcal{A}) \subseteq  \mathcal{E}(\mathcal{A})$. This relation is known as \emph{causal wedge inclusion}. Backgrounds  that do not satisfy causal wedge inclusion are not viable as holographic duals of a boundary field theory.

\subsection{Entanglement entropy of a disk in 4-dimensional ECG}
\label{subsec:EEdiskECG}
Consider Euclidean ${\rm AdS}_4$ written in boundary polar coordinates:
\begin{equation} \label{DiskECG:AdS_metric}
\mathrm{d}s^2 = \frac{L_{\star}^2}{z^2} \left( \mathrm{d}\tau^2 + \mathrm{d}z^2 + \mathrm{d}r^2 + r^2 \mathrm{d}\phi^2 \right) ~ .
\end{equation}
Take the boundary region to be defined as $r \leq R$, $\phi \in [0, 2 \pi)$ at $z = 0$ and some fixed $\tau$. This is a disk which will produce an entanglement surface penetrating into the bulk which we choose to parametrize as $r = \xi$, $z = Z(\xi)$ (plus the angular, symmetric direction). In Appendix \ref{app:GeometryDisk}, the different geometric quantities relevant for the computation of the entanglement entropy functional associated with this surface can be found. We will present in a moment the explicit form of the functionals following both the minimal and the non-minimal prescriptions, but notice a relevant fact that can already be anticipated at this point. For both funcionals \eqref{FunctionalCubic:functional_XD} and \eqref{FunctionalCubic:functional_JC} there is a term ($S_{R^2}$) which involves quadratic curvature contractions (proportional to $1/L_{\star}^4$ when evaluated on pure AdS, independently of the entanglement surface) and an extra piece ($S_{K^2 R} + S_{K^4}$) which depends on the prescription and which is at least quadratic in extrinsic curvatures. As shown at the end of Appendix \ref{app:GeometryDisk}, the RT or minimal area surface with correct boundary conditions is just an spherical shell $Z(\xi) = \sqrt{R^2 - \xi^2}$, which happens to have vanishing extrinsic curvature. This surface will therefore also minimize the complete functional (whether it is \eqref{FunctionalCubic:functional_XD}, \eqref{FunctionalCubic:functional_JC}, or in fact any other prescription), because the $S_{R^2}$ piece in pure AdS will be just a constant times the area of the surface, while the extra parts will produce terms in the Euler-Lagrange equations for extremization proportional to the extrinsic curvature, which therefore vanish for the spherical shell.

Just as a reassuring check, we can explicitly write the functionals obtained with each of the prescriptions. For the minimal one:
\begin{equation} \label{DiskECG:functional_minimal}
S^{min}_{disk} = \frac{\pi L_{\star}^2}{2 G_N} \int \mathrm{d} \xi \left[ \frac{\xi \sqrt{1 + (Z')^2}}{Z^2} + \frac{3 f_{\infty}^2 \mu}{4 \xi^2 Z^2 \left[ 1 + (Z')^2 \right]^{9/2}} \; \mathcal{S}^{min} \right] ~ , 
\end{equation}
where
\begin{align} \label{DiskECG:S_minimal}
\mathcal{S}^{min} \equiv & - 2 \xi^3 \left[3 - 2 (Z')^4 \right] \left[ 1 + (Z')^2 \right]^3 - 4 \xi^2 Z \left[ 1 + (Z')^2 \right]^2 \left[ 3 + 2 (Z')^2 \right] \left[ Z' + (Z')^3 + \xi Z'' \right] \nonumber \\
 & - \xi Z^2 \left[1 + (Z')^2 \right] \left[ 3 \left( Z' + (Z')^3 \right)^2 + 2 \xi Z' Z'' \left( 1 + (Z')^2 \right) \left( 7 + 4 (Z')^2 \right) + 3 \xi^2 (Z'')^2 \right] \nonumber \\
 & - Z^3 \left[ Z' + (Z')^3 + \xi Z'' \right]^3 - Z^4 Z' Z'' \left[ \left( Z' + (Z')^3 \right)^2 + \xi^2 (Z'')^2 \right] ~.
\end{align}
For the non-minimal prescription:
\begin{equation} \label{DiskECG:functional_non-minimal}
S^{non-min}_{disk} = \frac{\pi L_{\star}^2}{2 G_N} \int \mathrm{d} \xi \left[ \frac{\xi \sqrt{1 + (Z')^2}}{Z^2} + \frac{3 f_{\infty}^2 \mu}{8 \xi^2 Z^2 \left[ 1 + (Z')^2 \right]^{9/2}} \; \mathcal{S}^{min} \right] ~ ,
\end{equation}
where now
\begin{align} \label{DiskECG:S_non-minimal}
\mathcal{S}^{non-min} \equiv & - 2 \xi^3 \left[3 - 4 (Z')^4 \right] \left[ 1 + (Z')^2 \right]^3 - 4 \xi^2 Z \left[ 1 + (Z')^2 \right]^2 \left[ 3 + 4 (Z')^2 \right] \left[ Z' + (Z')^3 + \xi Z'' \right] \nonumber \\
 & - \xi Z^2 \left[1 + (Z')^2 \right] \left[ \left( Z' + (Z')^3 \right)^2 + 2 \xi Z' Z'' \left( 1 + (Z')^2 \right) \left( 1 + 8 (Z')^2 \right) + \xi^2 (Z'')^2 \right] \nonumber \\
 & - Z^3 \left[ Z' + (Z')^3 - 3 \xi Z'' \right] \left[ 3 Z' + 3 (Z')^3 - \xi Z'' \right] \left[ Z' + (Z')^3 + \xi Z'' \right] \nonumber \\
 & - Z^4 Z' Z'' \left[ 3 \left( Z' + (Z')^3 \right)^2 - 8 \xi Z'' \left( Z' + (Z')^3 \right) + 3 \xi^2 (Z'')^2 \right] ~.
\end{align}
While not immediate to discern from these complicated expressions for the functionals, one can compute their Euler-Lagrange equations and verify that $Z(\xi) = \sqrt{R^2 - \xi^2}$ is indeed a solution, as the general argument presented before shows.

Since the entanglement surface coincides with the minimal area one (\emph{i.e.}, the one we would have obtained without higher curvature corrections), and taking into account that the causal structure around pure AdS for ECG is also the same than for Einstein gravity, we conclude that causal wedge inclusion is respected in this case just like it is respected in pure Einstein gravity. Incidentally, this means that it is \emph{marginally} respected, which is to say that both the entanglement and the causal wedge coincide, as can be easily checked by verifyng that the projection of the causal wedge into the fixed $\tau$ slice coincides with the entanglement surface, $z^2 + r^2 = R^2$. The simple disk geometry in the boundary we have considered is therefore not able to constrain ECG in any way, since it respects the conditions imposed by subregion duality in AdS/CFT. Let us consider a different geometry to see whether we can extract any useful information.

\subsection{Entanglement entropy example: a strip in 4-dimensional ECG}
\label{subsec:EEexampleECG}

Similarly to what we did in the previous section, consider Euclidean $\rm{AdS}_4$ as the bulk metric, but written now in boundary Cartesian coordinates:
\begin{equation} \label{StripECG:AdS_metric}
\mathrm{d}s^2 = \frac{L_{\star}^2}{z^2} \left( \mathrm{d}\tau^2 + \mathrm{d}z^2 + \mathrm{d}x^2 + \mathrm{d}y^2 \right) ~ ,
\end{equation}
and take the boundary strip to be $x \in (-\ell/2, \ell/2)$, $y \in (- \infty, \infty)$ at $z = 0$ and some fixed $\tau$. We parametrize the entanglement surface with $(x,y)$ as $z = Z(x)$, since $y$ is a symmetry direction. We refer to Appendix \ref{app:GeometryStrip} for the computation of the relevant geometric quantities of this entanglement surface. Taking those results into account, we can write the functional for a strip in ECG \eqref{Review:Action} following both of the splitting prescriptions. With the minimal one we obtain, starting from \eqref{FunctionalCubic:functional_XD}:
\begin{equation} \label{StripECG:functional_minimal}
S^{min}_{strip} = \frac{L_{\star}^2}{4 G_N} \int \mathrm{d} y \, \mathrm{d} x \left[ \frac{\sqrt{1 + (Z')^2}}{Z^2} + \frac{3 f_{\infty}^2 \mu}{4 Z^2 \left[ 1 + (Z')^2 \right]^{9/2}} \; \mathcal{S}^{min} \right] ~ , 
\end{equation}
where
\begin{align} \label{StripECG:S_minimal}
\mathcal{S}^{min} \equiv & - 6 + 12 (Z')^8 + 4 (Z')^{10} - 12 Z Z'' - 3 Z^2 (Z'')^2 - Z^3 (Z'')^3 + (Z')^6 \left( 6 - 8 Z Z'' \right) - \nonumber \\
 & - 14 (Z')^4 \left( 1 + 2 Z Z'' \right) - (Z')^2 \left[ 18 + 32 Z Z'' + 3 Z^2 (Z'')^2 \right] ~.
\end{align}
We can start from \eqref{FunctionalCubic:functional_JC} instead to obtain the functional following the non-minimal prescription:
\begin{equation} \label{StripECG:functional_non-minimal}
S^{non-min}_{strip} = \frac{L_{\star}^2}{4 G_N} \int \mathrm{d} y \, \mathrm{d} x \left[ \frac{\sqrt{1 + (Z')^2}}{Z^2} + \frac{3 f_{\infty}^2 \mu}{8 Z^2 \left[ 1 + (Z')^2 \right]^{9/2}} \; \mathcal{S}^{non-min} \right] ~ ,
\end{equation}
where now
\begin{align} \label{StripECG:S_non-minimal}
\mathcal{S}^{non-min} \equiv & -6 + 24 (Z')^8 + 8 (Z')^{10} - 12 Z Z'' - Z^2 (Z'')^2 - 3 Z^3 (Z'')^3 - 2 (Z')^6 \left( -9 + 8 Z Z'' \right) - \nonumber \\
 & - 2 (Z')^4 \left( 5 + 22 Z Z'' \right) - (Z')^2 \left( 18 + 40 Z Z'' + Z^2 (Z'')^2 \right)  ~.
\end{align}
A couple of comments are in order. First of all, recall that in order to have AdS with curvature radius $L^2_\star = L^2 / f_{\infty}$ as a background in ECG, $f_{\infty}$ must satisfy:
\begin{equation} \label{StripECG:background_equation}
1 - f_{\infty} + \mu f_{\infty}^3 = 0 ~,
\end{equation}
and we are taking the solution of this equation which has positive effective Newton constant and connects with the GR solution $f_{\infty} = 1$ in the limit $\mu \to 0$, given by \eqref{ReviewECG:Finfinity}. Second, both functionals have an obvious IR divergence, since the $y$ integral is infinite. An IR-regulator must be therefore included, cutting the strip at some fixed length. Notice also another suprising feature of the previous functionals. As already mentioned, the entanglement entropy of the strip should be given by the functional obtained employing the \emph{correct} splitting prescription for ECG (which might be none of the previous ones), and evaluated at the surface determined by the $Z(x)$ which extremizes that functional. If the correct splitting happens to be one of the previous ones, then we have to obtain the Euler-Lagrange equation for $Z(x)$ starting from the corresponding functional. In both cases, we have up to second derivatives of $Z$, so the corresponding differential equation for $Z$ will be fourth order. This contrasts with the second-order character of the equations of motion for the perturbations around a maximally symmetric background in ECG, which is one of its defining properties.

In the following section we disregard this fact and assume each of the functionals to be correct, as a test to see what would happen. We numerically solve for the surface profile $Z(x)$ which extremizes the corresponding functional,\footnote{We also had to numerically solve for $X(z)$ in order to construct the whole surface profile, as explained in Appendix \ref{app:DetailsNumerical}.} and we plot the result. This will serve as a probe to understand how the higher-derivative corrections to the entanglement entropy change the bulk surface in which the functional is to be evaluated.

\subsubsection{Numerical results} 
\label{subsec:NumericalResults}

\begin{figure}[ht!]

\begin{subfigure}{0.5\textwidth}
\includegraphics[width=0.9\linewidth, height=5cm]{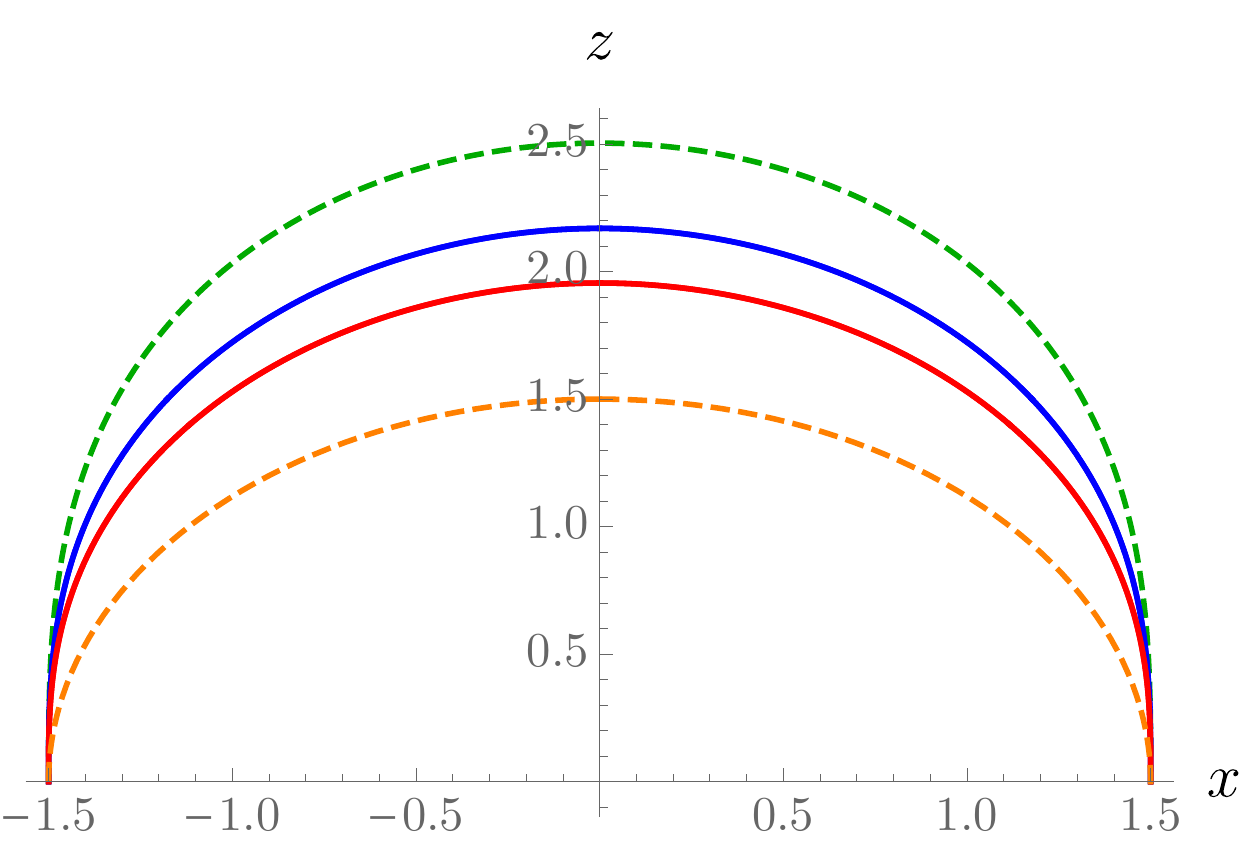}
\caption{$\mu = -0.50$}
\label{FigNR:Mu-0_50}
\end{subfigure}
\begin{subfigure}{0.5\textwidth}
\includegraphics[width=0.9\linewidth, height=5cm]{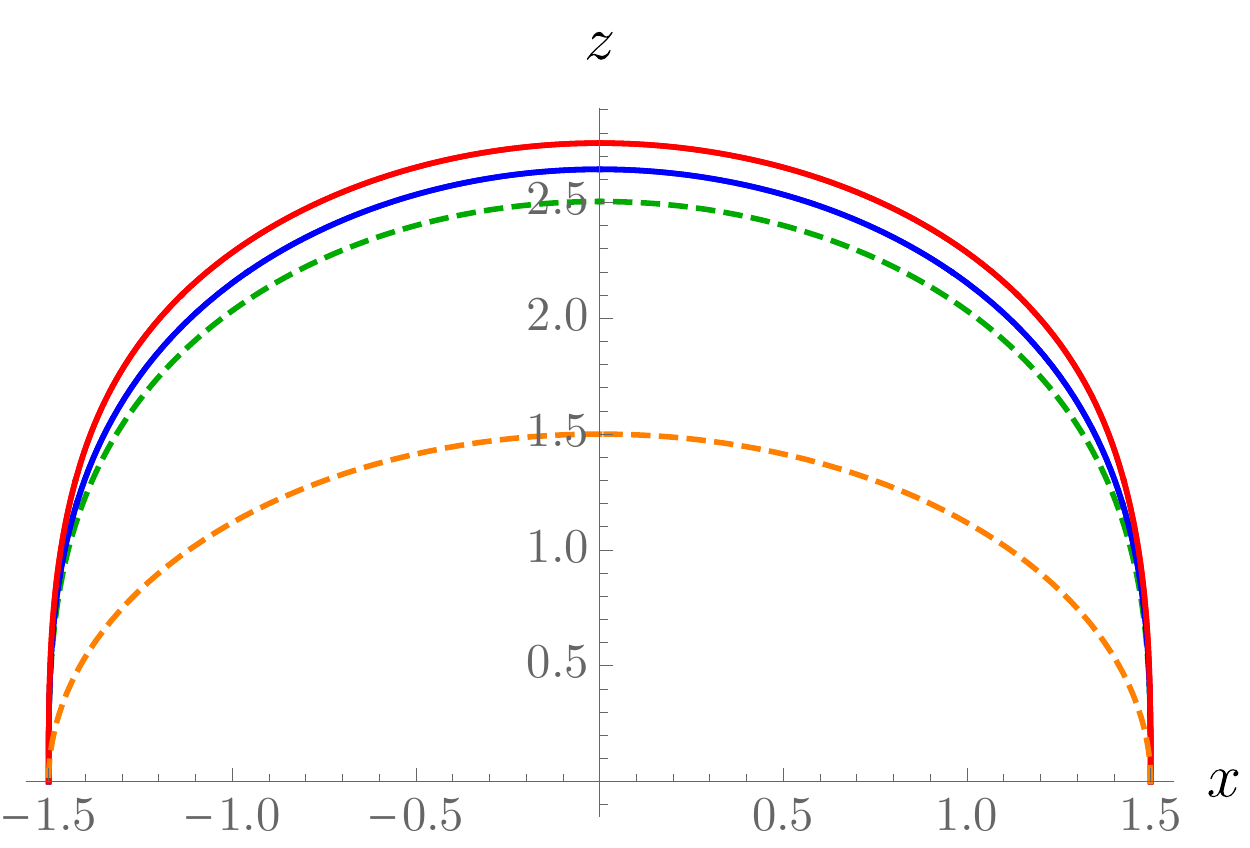}
\caption{$\mu = +0.010$}
\label{FigNR:XDMu+0_10}
\end{subfigure}

\caption{Causal wedge (orange, dashed) and entanglement surfaces corresponding to ECG in the case of minimal prescription (blue), ECG in the case of non-minimal prescription (red), and Einstein gravity (green, dashed) \cite{HubenySurfaces} for the boundary strip length $\ell =3$ and two different values of $\mu$ outside the interval $-0.00322 \leq \mu \leq 0.00312$. In both cases, we verify that causal wedge inclusion is satisfied for both prescriptions.}
\label{FigNR:OtherMu}
\end{figure}

\begin{figure}[ht!]

\begin{subfigure}{0.5\textwidth}
\includegraphics[width=0.85\linewidth, height=4.5cm]{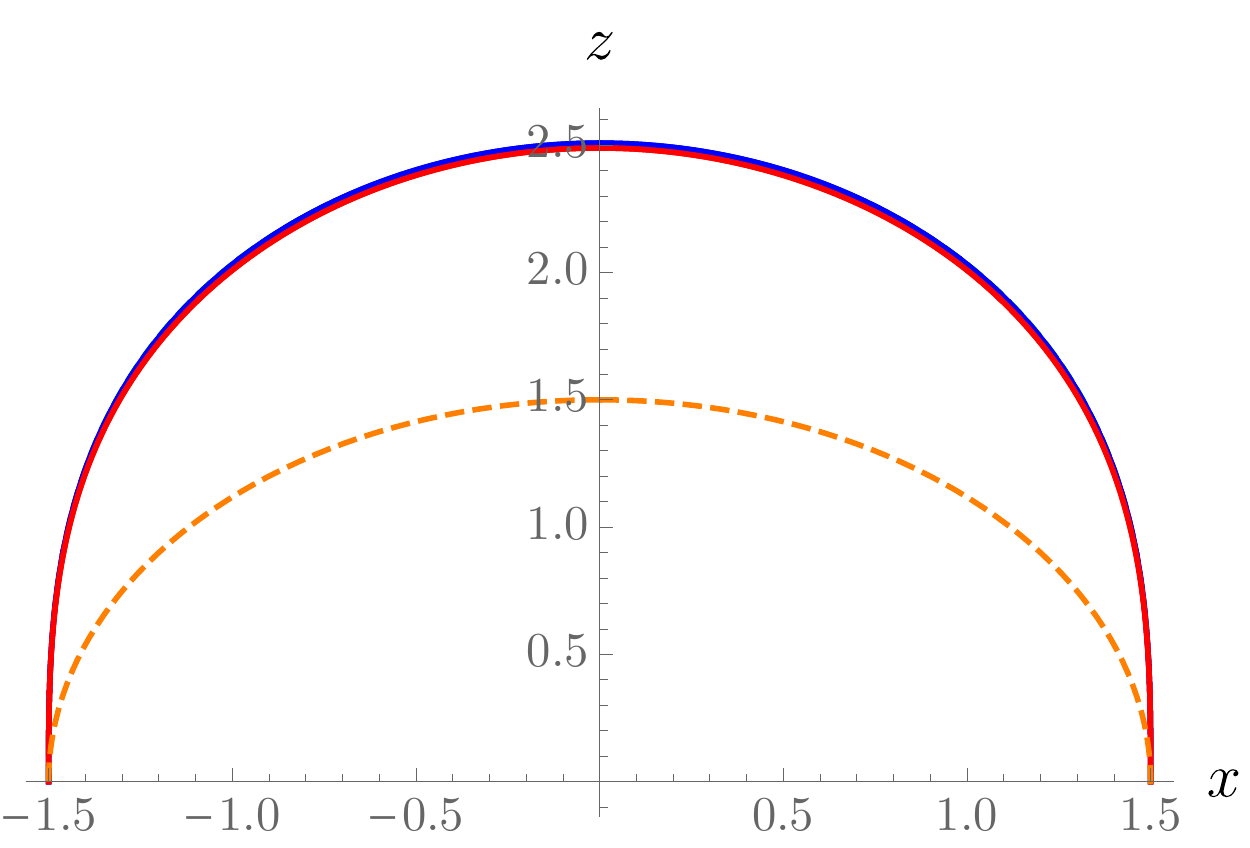} 
\caption{\footnotesize{$\mu=-0.002$}}
\label{FigNR:Mu-0_002}
\end{subfigure}
\begin{subfigure}{0.5\textwidth}
\includegraphics[width=0.85\linewidth, height=4.5cm]{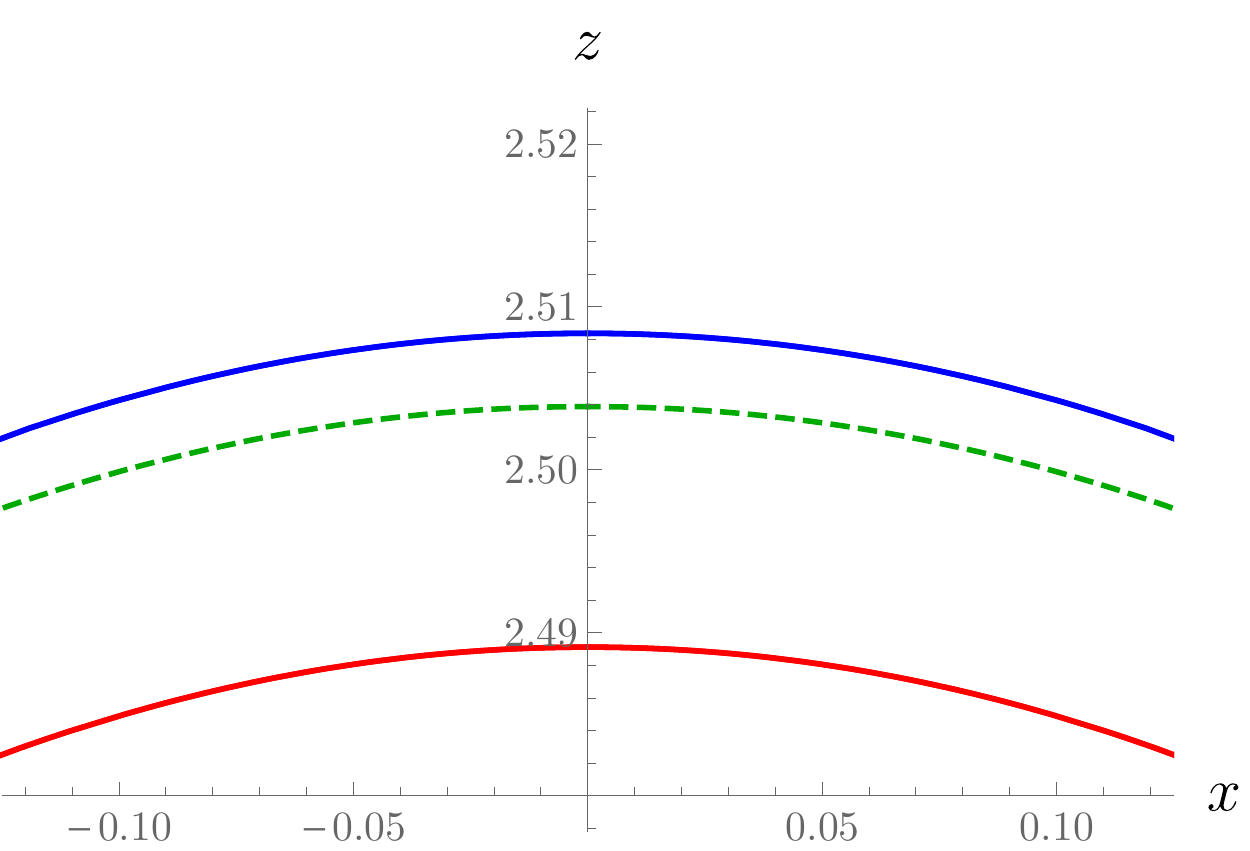}
\caption{\footnotesize{$\mu=-0.002$}}
\label{FigNR:Mu-0_002Zoomin}
\end{subfigure}

\caption{\footnotesize{Causal wedge (orange, dashed) and entanglement surfaces corresponding to ECG in the case of minimal prescription (blue), ECG in the case of non-minimal prescription (red), and Einstein gravity (green, dashed) \cite{HubenySurfaces} for $\ell = 3$ and $\mu=-0.002$, a value within the interval $-0.00322 \leq \mu \leq 0.00312$. We include a close-up of the entanglement surfaces near $x=0$ to tell them apart better.}}
\label{FigNR:SmallerIntervalMu}
\end{figure}

In this section we present some of the curves for the numerical solutions obtained by minimizing the functionals \eqref{StripECG:functional_minimal} and \eqref{StripECG:functional_non-minimal}. The boundary region is a strip of width  $\ell=3$ in the $x$ direction and infinite in the $y$ direction. We solved the fourth-order equations \eqref{MinimizationFunctionals:XDeqZ} and \eqref{MinimizationFunctionals:RXMeqZ} to obtain the corresponding entanglement wedge. All the details regarding the numerical procedure, along with more cases of $\mu$ values for which the solution was obtained, are presented in Appendix \ref{app:DetailsNumerical}. The resulting plots, in which we also include the causal wedge, can be found in Figures \ref{FigNR:OtherMu} and \ref{FigNR:SmallerIntervalMu}.

These plots show that the causal wedge is safely included in the entanglement wedge for both prescriptions, and for all values of the coupling within the range tested (in Appendix \ref{app:DetailsNumerical} we present examples which prove this to be the case for, at least, $-10^4 \leq \mu \leq + 0.01$). Thus, just like for the boundary disk, causal wedge inclusion applied to the boundary strip in Poincar\'e AdS does not constrain the allowed values of the coupling in ECG. Note that causal wedge inclusion in a Gauss-Bonnet black hole background does constrain the allowed values of the coupling \cite{Caceres:2019pok}. A crucial ingredient to arrive at the results of \cite{Caceres:2019pok} was the modified causal structure of the higher curvature theory due to the presence of superluminal modes. Here we are working in AdS and the causal structure is still governed by null rays. In addition, the disk has the same entanglement surface in both Einstein gravity and ECG, and therefore although causal wedge inclusion is marginally respected, it is so for all values of the coupling. For the strip geometry, in the case of Einstein gravity ($\mu = 0$), causal wedge inclusion is safely respected, as the previous plots show. Thus, to violate causal wedge inclusion, it would be necessary a big modification of the surface due to turning on the coupling $\mu$. This does not happen.

Clearly, to explore the effects of causal wedge inclusion in ECG it would be better to study a black hole background, just like the authors did in \cite{Caceres:2019pok} in Gauss-Bonnet gravity. However, in order to construct the causal wedge in a higher curvature theory, a complete investigation of the superluminal modes that determine the causal structure is required. A different possible avenue to harness the power of causal wedge inclusion is to investigate perturbative modifications of the disk region and solve the Euler-Lagrange equations derived from \eqref{DiskECG:functional_minimal} or \eqref{DiskECG:functional_non-minimal} with the perturbed boundary condition, but still in pure AdS, similarly to what \cite{Mezei:2014zla} did for Einstein gravity. Since causal wedge inclusion is marginally respected there, it could happen that a perturbative violation occurs for some values of the coupling, which would therefore have to be excluded. The complicated form of the Euler-Lagrange equations makes this an extremely involved problem at the technical level, which we therefore leave as an open question.

\section{Conclusions and future directions}
\label{sec:conclusions}

In the present paper we have obtained the entanglement entropy functional for a generic cubic gravitational theory not involving derivatives of the curvature tensor in the action. This constitutes our main result. We have done this using two different prescriptions: the minimal one, introduced in \cite{Dong:entanglement_entropy}, and the non-minimal one, presented in \cite{MiaoSplitting2,CampsBoundaryCones} and which is known to be perturbatively valid. The existence of these two alternative functionals is an explicit manifestation of the splitting problem one is forced to face when deriving the entanglement entropy functional for cubic or higher order gravitational theories. Despite knowing that for a particular theory there must be a single, correct functional, we have also performed some consistency checks for both of them in a particular cubic theory. In particular, we  investigated whether the functionals obtained for Einsteinian Cubic Gravity produce via minimization entanglement surfaces which satisfy the causal wedge inclusion property for both a boundary disk and a boundary strip in Poincar\'e AdS. We find that, for all the values of the couplings studied, causal wedge inclusion is respected. 

Our work makes it possible to investigate several questions that will advance our understanding of the role cubic theories play in a holographic context.

\begin{itemize}
	\item[]\emph{Bit threads}\\
In \cite{Freedman:2016zud} the authors put forward an alternative formulation of the holographic entanglement entropy that does not rely on minimizing an area functional but invokes a divergenceless vector field, dubbed bit threads. Many aspects of bit threads have been studied \cite{Agon:2019qgh,Agon:2020mvu,Du:2019emy,Harper:2019lff,Bao:2019wcf,Headrick:2017ucz}, but a formulation of bit threads for higher-derivative theories was missing. Recently this gap was closed in \cite{Harper:2018sdd}, where the authors derived a bit thread formulation for a general higher-derivative theory. Now that we have the entanglement functional for cubic theories in the minimal and non-minimal splitting, it would be interesting to investigate if using bit threads we can understand better the splitting problem. 
	\item[]\emph{Dynamics}\\
Holographic entanglement entropy in dynamical backgrounds  has been widely studied in Einstein gravity, with and without charge. It led to interesting insights regarding the thermalization time \cite{Balasubramanian:2011ur,Caceres:2012em,Galante:2012pv}. Similar work has been carried out for Gauss-Bonnet backgrounds \cite{Zeng:2013mca, Li:2013cja,  Caceres:2015bkr}. If a Vaidya type of solution can be written down for black holes in cubic gravity, then dynamical studies in these theories along the lines suggested would be quite interesting. 
	\item[]\emph{More general boundary regions}\\
We have learned valuable lessons studying HEE for different boundary regions: as an example, the divergence structure of the holographic entanglement entropy on regions with corners uncovered cut-off independent coefficients. These coefficients were shown to be universal and to encode important field theory data in Einstein and Gauss-Bonnet theories \cite{Myers_2012,Bueno_2015}. It would be interesting to study regions with corners in cubic theories. 
	\item[] \emph{Derive the correct splitting for finite coupling} \\
Clearly, an important open question is to formally find the correct splitting for a general cubic theory with finite coupling. This highly technical problem can be approached following \cite{MiaoSplitting2,CampsBoundaryCones}. 
	\item []\emph{Causal structure of cubic theories}\\
In section \ref{sec:StripECG}, as an example of the functional we derived, we calculated the entanglement surfaces of a disk and a strip regions in an AdS background solution of ECG. More interesting phenomena regarding the causal and entanglement wedges can be expected if we were to consider black holes in any cubic gravity theory. However, black hole backgrounds in higher-derivative theories will generically have superluminal modes, and their causal structure is no longer determined by null rays. A complete study of  the causal structure and hyperbolicity of equations of motion of cubic theories similar to what was done for Gauss Bonnet \cite{Reall:2014pwa,Izumi:2014loa,Takahashi_2010} is of interest not only for the GR communities but also in a holographic context.
	\item[]\emph{Perturbative calculations in field theory} \\
The non-minimal functional is known to be the correct one for any perturbative cubic theory which does not include covariant derivatives of the curvature tensors. Furthermore, perturbatively, we know that the entanglement entropy functional can be evaluated in the Ryu-Takayanagi surface obtained in General Relativity. This is because the area term of the functional is minimal for the Ryu-Takayanagi surface, and therefore even if the surface changes at first order, it does not produce a change of that part of the functional. The first order variation of the Ryu-Takayanagi surface can also be neglected for the remaining part of the functional coming from the higher-derivative terms. This approach will be explored elsewhere \cite{BuenoVilar}.
\end{itemize}

\acknowledgments{ We are deeply indebted to Joan Camps for insightful  explanations and correspondence. It is a pleasure to thank also  Pablo Bueno,  Jos\'e Edelstein  and Julio Oliva for useful discussions. EC and RC are supported by the National Science Foundation (NSF) Grant No. PHY1820712. RC is also supported  by NSF  Grant No. PHY-1620610. The work of AVL is supported by the Spanish MECD fellowship FPU16/06675, and by MINECO FPA2017-84436-P, Xunta de Galicia ED431C 2017/07, Xunta de Galicia (Centro singular de investigaci\'on de Galicia accreditation 2019-2022) and the European Union (European Regional Development Fund -- ERDF), ``Mar\'\i a de Maeztu'' Units of Excellence MDM-2016-0692, and the Spanish Research State Agency. AVL is also pleased to thank the University of Texas at Austin, where part of this work was done, for their warm hospitality.}
 
\appendix
\section{Calculation of the entanglement functional}\label{app:details_func}

\subsection{A first, detailed example}

In order to understand better how the entanglement entropy functional is obtained for a generic cubic theory, let us present an example in detail. Consider the following (Euclidean) Lagrangian:
\begin{equation} \label{details_func:example_Lagrangian}
\mathcal{L}_E = \lambda R_{\mu \nu \rho \sigma} R^{\mu \nu \rho \sigma} R ~ ,
\end{equation}
where $\lambda$ is a constant. We will compute separately each of the terms in the general form of the functional \eqref{SplittingProblem:general_functional}, and for the second one we will do it following the two prescriptions presented in section \ref{sec:HHEinHDG}. First of all, for the Wald term, we need the following derivative:
\begin{equation} \label{details_func:example_Wald}
\frac{\partial \mathcal{L}_E}{\partial R_{z \bz z \bz}} = -2 \lambda R_{\mu \nu \rho \sigma} R^{\mu \nu \rho \sigma} + 2 \lambda R R^{z \bz z \bz} ~ .
\end{equation}
If we want to use this expression in a situation in which we do not have at our disposal the set of (complex) coordinates adapted to the surface, as will generically be the case, we need to covariantize the last term. This is done by going to the orthonormal basis \eqref{SplittingProblem:orthonormal_vectors}:
\begin{equation} \label{details_func:orthonormal_vectors}
n_1 = \sqrt{\frac{z}{\bz}} \partial_z + \sqrt{\frac{\bz}{z}} \partial_{\bz}  ~, \quad n_2 = i \left( \sqrt{\frac{z}{\bz}} \partial_z - \sqrt{\frac{\bz}{z}} \partial_{\bz} \right) ~ .
\end{equation}
The simplest way to do this is by first writing the expression as a contraction in the complex coordinates $z$ and $\bz$, using the fact that the only non-vanishing components of the metric are $G_{z \bz} = G_{\bz z} = 1/2$:
\begin{equation}
R^{z \bz z \bz} = - 4 R^{z \bz}{}_{z \bz} = - 2 R^{\alpha \beta}{}_{\alpha \beta} ~ ,
\end{equation}
where the antisymmetry of the Riemann tensor in the two pairs of indices has been taken into account. Now we use the fact that a contraction in an $\alpha$-index can be substituted for a contraction in orthonormal directions, since tangent directions have no $z$ or $\bz$ components:
\begin{equation}
T^{\alpha}{}_{\alpha} = T^a{}_b (n_a)^{\alpha} (n^b)_{\alpha} = T^a{}_a ~ ,
\end{equation}
where $T$ is any tensor, possibly containing extra indices. The Wald term is then:
\begin{equation} \label{details_func:example_Wald_simplified}
\frac{\partial \mathcal{L}_E}{\partial R_{z \bz z \bz}} = -2  \lambda R_{\mu \nu \rho \sigma} R^{\mu \nu \rho \sigma} - 4 \lambda R R^{a b}{}_{a b} ~ .
\end{equation}
This is written in a form which can be evaluated in any set of coordinates just constructing the two orthonormal vectors $n_a$ to the surface.

Consider now the anomaly term. The first step is to obtain the second derivative of the Lagrangian:
\begin{equation} \label{details_func:second_derivative_L}
\frac{\partial^2 \mathcal{L}_E}{\partial R_{z i z j} \partial R_{\bz k \bz l}} = 2 \lambda g^{i(k} g^{l)j} R ~ .
\end{equation}
Now we have to expand $R$ following the two prescriptions. For this we have to write it in terms of Riemann tensor components and expand them following \eqref{SplittingProblem:expansions_Riemann}:
\begin{align} \label{details_func:Ricci_expanded}
R & = G^{\alpha \beta} R_{\alpha \beta} + g^{ij} R_{ij} = G^{\alpha \beta} G^{\gamma \delta} R_{\alpha \gamma \beta \delta} + 2 G^{\alpha \beta} g^{ij} R_{\alpha i \beta j} + g^{ij} g^{kl} R_{i k j l} = \\
\nonumber & = G^{\alpha \beta} G^{\gamma \delta} R_{\alpha \gamma \beta \delta} + 2 G^{\alpha \beta} g^{ij} \tilde{R}_{\alpha i \beta j} - 2 G^{\alpha \beta} g^{ij} Q_{\alpha \beta i j} + g^{ij} g^{kl} r_{i k j l} + 3 K_{\alpha i j} K^{\alpha i j} - K_{\alpha} K^{\alpha} ~ .
\end{align}
Now, in the minimal prescription the last two terms have $q_A = 1$, while all the rest have $q_A = 0$ (notice that $G^{\alpha \beta} Q_{\alpha \beta i j}$ involves only $Q_{z \bz i j}$). Therefore, when doing the $A$ expansion we have to multiply them by $1/2$. Doing that and then rewriting back everything in terms of the Ricci scalar, the minimal prescription gives:
\begin{equation} \label{details_func:Ricci_minimal}
\sum_A \left( \frac{R}{1 + q_A} \right)_A^{min} = R - \frac{3}{2} K_{\alpha i j} K^{\alpha i j} + \frac{1}{2} K_{\alpha} K^{\alpha} ~ .
\end{equation}
For the non-minimal prescription we need to rewrite $R_{z \bz z \bz}$ and $Q_{z \bz i j}$ in terms of the primed versions \eqref{SplittingProblem:primed_tensors}. This is done as follows:
\begin{align}
G^{\alpha \beta} G^{\gamma \delta} R_{\alpha \gamma \beta \delta} & = - 8 R_{z \bz z \bz} = G^{\alpha \beta} G^{\gamma \delta} R'_{\alpha \gamma \beta \delta} + K_{\alpha i j} K^{\alpha i j} - K_{\alpha} K^{\alpha} ~ , \\
G^{\alpha \beta} g^{ij} Q_{\alpha \beta i j} & = 4 g^{ij} Q_{z \bz i j} = G^{\alpha \beta} g^{ij} Q'_{\alpha \beta i j} + 2 K_{\alpha i j} K^{\alpha i j} - K_{\alpha} K^{\alpha} ~ .
\end{align}
The Ricci scalar is now written in terms of the basic objects for this prescription as:
\begin{equation}
R = G^{\alpha \beta} G^{\gamma \delta} R'_{\alpha \gamma \beta \delta} + 2 G^{\alpha \beta} g^{ij} \tilde{R}_{\alpha i \beta j} - 2 G^{\alpha \beta} g^{ij} Q'_{\alpha \beta i j} + g^{ij} g^{kl} r_{i k j l} ~ ,
\end{equation}
so that there are no terms with non-vanishing $q_A$. We can then immediately rewrite everything in terms of the Ricci scalar, obtaining:
\begin{equation} \label{details_func:Ricci_non-minimal}
\sum_A \left( \frac{R}{1 + q_A} \right)_A^{non-min} = R ~ .
\end{equation}
This completes the $A$ expansion for the non-minimal prescription. Notice that, in \eqref{details_func:second_derivative_L}, the metric tensors are not affected by the expansion, and we can contract them with the extrinsic curvatures appearing in the general formula \eqref{SplittingProblem:general_functional} as follows:
\begin{equation}
g^{i(k} g^{l)j} K_{z i j} K_{\bz k l} = K_{z i j} K_{\bz}{}^{i j} = \frac{1}{4} K_{\alpha i j} K^{\alpha i j} ~ .
\end{equation}
All contractions in $\alpha$ indices can now be traded for contractions in $a$ indices as explained when discussing the Wald term.

We conclude this little example by collecting all contributions, which would produce the following entanglement entropy functionals for the Lagrangian \eqref{details_func:example_Lagrangian}:
\begin{align}
S_{EE}^{min} & = 4 \pi \lambda \int d^{D-2} y \sqrt{g} \left[ - R_{\mu \nu \rho \sigma} R^{\mu \nu \rho \sigma} - 2 R R^{a b}{}_{a b} + 2 K_{a i j} K^{a i j} R \right. - \nonumber \\ 
 & \qquad \qquad \qquad \qquad \qquad \left. - 3 K_{a i j} K^{a i j} K_{b k l} K^{b k l} + K_{a i j} K^{a i j} K_{b} K^{b} \right]  ~ , \\
S_{EE}^{non-min} & = 4 \pi \lambda \int d^{D-2} y \sqrt{g} \left[ - R_{\mu \nu \rho \sigma} R^{\mu \nu \rho \sigma} - 2 R R^{a b}{}_{a b} + 2 K_{a i j} K^{a i j} R \right] ~ .
\end{align}
The following sections contain the contributions to both the Wald and anomaly terms of the functional using both prescriptions. We include all numerical factors except for the global $2 \pi$ appearing in \eqref{SplittingProblem:general_functional}.

\subsection{Minimal prescription}
\begin{flalign} 
 & \mathcal{L}_{3,8} = R^3 ~ , && \nonumber\\
 & \text{Wald}: - 6 R^2 ~ , &&\label{eq:func_XD_8} \\
 & \text{Anomaly}: 0 ~ . && \\ \cline{1-2}
 \label{functional_cubic:lagrangian7_derivativesXD}
 & \mathcal{L}_{3,7} = R_{\mu \nu} R^{\mu \nu} R ~ , && \nonumber\\
 & \text{Wald}: - 2 R_{\mu \nu} R^{\mu \nu} - 2 R_a{}^a R ~ , && \\
 & \text{Anomaly}: K_a K^a \left( R - \frac{3}{2} K_{bij} K^{bij} + \frac{1}{2} K_b K^b \right) ~ . && \\\cline{1-2}
 \label{functional_cubic:lagrangian6_derivativesXD}
 & \mathcal{L}_{3,6} = R_{\mu}{}^{\nu} R_{\nu}{}^{\rho} R_{\rho}{}^{\mu} ~ , && \nonumber\\
 & \text{Wald}: - 3 R_{a \mu} R^{a \mu} ~ , && \\
 & \text{Anomaly}: \frac{3}{2} K_a K^a \left( R_b{}^b - \frac{1}{2} K_{bij} K^{bij} \right) ~ . && \\\cline{1-2}
 \label{functional_cubic:lagrangian5_derivativesXD}
 & \mathcal{L}_{3,5} = R^{\mu \rho} R^{\nu \sigma} R_{\mu \nu \rho \sigma} ~ , && \nonumber\\
 & \text{Wald}: - 2 R_{\mu \nu} R_{a}{}^{\mu a \nu} + R_{ab} R^{ab} - R_a{}^a R_b{}^b ~ , && \\
 & \text{Anomaly}: 2 K_a K^{a}{}_{ij} \left( R^{ij} - K_{b}{}^{i}{}_{m} K^{bjm} + \frac{1}{2} K_{b} K^{bij} \right) - \frac{1}{2} K_a K^a R_{bc}{}^{bc} ~ . && \\\cline{1-2}
 \label{functional_cubic:lagrangian4_derivativesXD}
 & \mathcal{L}_{3,4} = R_{\mu \nu \rho \sigma} R^{\mu \nu \rho \sigma} R ~ , &&\nonumber \\
 & \text{Wald}: - 2 R_{\mu \nu \rho \sigma} R^{\mu \nu \rho \sigma} - 4 R R_{ab}{}^{ab} ~ , && \\
 & \text{Anomaly}: 4 K_{a i j} K^{a i j} \left( R - \frac{3}{2} K_{b k l} K^{b k l} + \frac{1}{2} K_b K^b \right) ~ . && \\\cline{1-2}
 \label{functional_cubic:lagrangian3_derivativesXD}
 & \mathcal{L}_{3,3} = R^{\mu \nu \rho}{}_{\sigma} R_{\mu \nu \rho \tau} R^{\sigma \tau} ~ , &&\nonumber \\
 & \text{Wald}: - R_{a \mu \nu \rho} R^{a \mu \nu \rho} - 4 R_{a \mu} R_b{}^{a b \mu} ~ , && \\
 & \text{Anomaly}: 2 K_{a i k} K^a{}_{j}{}^{k} R^{ij} + K_{aij} K^{aij} R_b{}^b + 2 K_a K^{a}{}_{ij} R_{b}{}^{ibj} - && \\
 & \qquad \qquad \qquad - 2 K_{a i}{}^j K^a{}_j{}^k K_{b k}{}^l K^b{}_l{}^i - \frac{1}{2} K_{aij} K^{aij} K_{bkl} K^{bkl} ~ . && \nonumber \\ \cline{1-2}
 \label{functional_cubic:lagrangian2_derivativesXD}
 & \mathcal{L}_{3,2} = R^{\mu \nu}{}_{\rho \sigma} R^{\rho \sigma}{}_{\lambda \tau} R^{\lambda \tau}{}_{\mu \nu} ~ , &&\nonumber \\
 & \text{Wald}: - 6 R_{a b \mu \nu} R^{a b \mu \nu} ~ , && \\
 & \text{Anomaly}: 12 K_{aik} K_{b j}{}^{k} R^{a b i j} + 12 K_{a i k} K^a{}_{j}{}^k R_{b}{}^{i b j} - 6 K_{a i}{}^j K_{bj}{}^k K^a{}_{k}{}^l K^b{}_l{}^i  ~ . && \\ \cline{1-2}
 \label{functional_cubic:lagrangian1_derivativesXD}
 & \mathcal{L}_{3,1} = R_{\mu}{}^{\rho}{}_{\nu}{}^{\sigma} R_{\rho}{}^{\lambda}{}_{\sigma}{}^{\tau} R_{\lambda}{}^{\mu}{}_{\tau}{}^{\nu} ~ , &&\nonumber \\
 & \text{Wald}: -3 \left( R_{a \mu}{}^{a}{}_{\nu} R_{b}{}^{\mu b \nu} - R_{a \mu b \nu} R^{a \nu b \mu} \right) ~ , && \\
 & \text{Anomaly}: 3 K_{a i j} K^{a}{}_{kl} R^{ikjl} + 6 K_{a i k} K_{b j}{}^{k} R^{a b i j} - \frac{3}{2} K_{a i j} K^{a i j} R_{bc}{}^{bc}  + && \label{eq:func_XD_1}\\
 & \qquad \qquad \quad + \frac{3}{2} K_{aij} K^{aij} K_{bkl} K^{bkl} - \frac{9}{2} K_{a i}{}^j K_{bj}{}^k K^a{}_{k}{}^l K^b{}_l{}^i + 3 K_{a i}{}^j K^a{}_j{}^k K_{b k}{}^l K^b{}_l{}^i ~ . && \nonumber
 \end{flalign}

\subsection{Non-minimal prescription}
\begin{flalign} 
 & \mathcal{L}_{3,8} = R^3 ~ , && \\
 & \text{Wald}: - 6 R^2 ~ , && \label{eq:func_JC_8}\\
 & \text{Anomaly}: 0 ~ . && \\ \cline{1-2}
 \label{functional_cubic:lagrangian7_derivativesJC}
 & \mathcal{L}_{3,7} = R_{\mu \nu} R^{\mu \nu} R ~ , && \\
 & \text{Wald}: - 2 R_{\mu \nu} R^{\mu \nu} - 2 R_a{}^a R ~ , && \\
 & \text{Anomaly}: K_a K^a R ~ . && \\\cline{1-2}
 \label{functional_cubic:lagrangian6_derivativesJC}
 & \mathcal{L}_{3,6} = R_{\mu}{}^{\nu} R_{\nu}{}^{\rho} R_{\rho}{}^{\mu} ~ , && \\
 & \text{Wald}: - 3 R_{a \mu} R^{a \mu} ~ , && \\
 & \text{Anomaly}: \frac{3}{2} K_a K^a R_b{}^b ~ . && \\\cline{1-2}
 \label{functional_cubic:lagrangian5_derivativesJC}
 & \mathcal{L}_{3,5} = R^{\mu \rho} R^{\nu \sigma} R_{\mu \nu \rho \sigma} ~ , && \\
 & \text{Wald}: - 2 R_{\mu \nu} R_{a}{}^{\mu a \nu} + R_{ab} R^{ab} - R_a{}^a R_b{}^b ~ , && \\
 & \text{Anomaly}: 2 K_a K^{a}{}_{ij} R^{ij} - \frac{1}{2} K_a K^a \left( R_{bc}{}^{bc} + \frac{1}{2} K_b K^b - \frac{1}{2} K_{b i j} K^{b i j} \right) ~ . && \\\cline{1-2}
 \label{functional_cubic:lagrangian4_derivativesJC}
 & \mathcal{L}_{3,4} = R_{\mu \nu \rho \sigma} R^{\mu \nu \rho \sigma} R ~ , && \\
 & \text{Wald}: - 2 R_{\mu \nu \rho \sigma} R^{\mu \nu \rho \sigma} - 4 R R_{ab}{}^{ab} ~ , && \\
 & \text{Anomaly}: 4 K_{a i j} K^{a i j} R ~ . && \\\cline{1-2}
 \label{functional_cubic:lagrangian3_derivativesJC}
 & \mathcal{L}_{3,3} = R^{\mu \nu \rho}{}_{\sigma} R_{\mu \nu \rho \tau} R^{\sigma \tau} ~ , && \\
 & \text{Wald}: - R_{a \mu \nu \rho} R^{a \mu \nu \rho} - 4 R_{a \mu} R_b{}^{a b \mu} ~ , && \\
 & \text{Anomaly}: 2 K_{aik} K^{a}{}_{j}{}^{k} R^{ij} + K_{aij} K^{aij} R_b{}^b + 2 K_a K^{a}{}_{i j} R_{b}{}^{i b j} + && \\
 & \qquad \qquad \qquad + K_a K^{a}{}_{i j} K_{b}{}^{i}{}_{k} K^{b j k} - K_a K^a{}_{i j} K_b K^{b i j} ~ . && \nonumber \\ \cline{1-2}
 \label{functional_cubic:lagrangian2_derivativesJC}
 & \mathcal{L}_{3,2} = R^{\mu \nu}{}_{\rho \sigma} R^{\rho \sigma}{}_{\lambda \tau} R^{\lambda \tau}{}_{\mu \nu} ~ , && \\
 & \text{Wald}: - 6 R_{a b \mu \nu} R^{a b \mu \nu} ~ , && \\
 & \text{Anomaly}: 12 K_{a i k} K_{b j}{}^{k} R^{a b i j} + 12 K_{a i k} K^{a}{}_{j}{}^{k} R_{b}{}^{ibj} - && \\
 & \qquad \qquad \qquad - 6 K_a K^{a}{}_{i j} K_{b}{}^{i}{}_{k} K^{b j k} - 6 K_{a i}{}^j K_{bj}{}^k K^a{}_{k}{}^l K^b{}_l{}^i + 12 K_{a i}{}^j K^a{}_{j}{}^k K_{b k}{}^l K^b{}_l{}^i  ~ . && \nonumber \\ \cline{1-2}
 \label{functional_cubic:lagrangian1_derivativesJC}
 & \mathcal{L}_{3,1} = R_{\mu}{}^{\rho}{}_{\nu}{}^{\sigma} R_{\rho}{}^{\lambda}{}_{\sigma}{}^{\tau} R_{\lambda}{}^{\mu}{}_{\tau}{}^{\nu} ~ , && \\
 & \text{Wald}: -3 \left( R_{a \mu}{}^{a}{}_{\nu} R_{b}{}^{\mu b \nu} - R_{a \mu b \nu} R^{a \nu b \mu} \right) ~ , && \\
 & \text{Anomaly}: 3 K_{a i j} K^{a}{}_{kl} R^{ikjl} + 6 K_{a i k} K_{b j}{}^{k} R^{a b i j} - \frac{3}{2} K_{a i j} K^{a i j} \left( R_{bc}{}^{bc} + \frac{1}{2} K_b K^b \right) + && \label{eq:func_JC_1}\\
 & + \frac{3}{4} K_{a i j} K^{a i j} K_{b k l} K^{b k l} + \frac{3}{2} K_{aij} K_{b k l} K^{bij} K^{akl} - \frac{9}{2} K_{a i}{}^j K_{bj}{}^k K^a{}_{k}{}^l K^b{}_l{}^i + 3 K_{a i}{}^j K^a{}_{j}{}^k K_{b k}{}^l K^b{}_l{}^i ~ . && \nonumber
 \end{flalign}
%

\section{Geometry of the entanglement surface for a boundary disk}
\label{app:GeometryDisk}

Let us collect here the results needed to evaluate the tensors appearing in the entanglement entropy functional for a boundary ball in Poincar\'e AdS (we consider the case of a 2-dimensional disk in the main body of the paper, but results will be presented here for the case of a general $(D-2)$-ball). Consider then the bulk metric:
\begin{equation} \label{GeometryDisk:AdSMetric}
\mathrm{d}s^2 = \frac{L_{\star}^2}{z^2} \left( \mathrm{d}\tau^2 + \mathrm{d}z^2 + \mathrm{d}r^2 + r^2 \mathrm{d}\Omega^2_{D-3}\right) ~ ,
\end{equation}
where the length scale $L_{\star}$ is determined by imposing this to be a solution of the equations of motion for the theory at hand. The ball in the boundary will be parametrized at fixed $\tau$ as $r \leq R$, so the surface anchored at its boundary and going into the bulk is parametrized as:
\begin{equation} \label{GeometryDisk:AdSBall}
\tau = \tau_0 ~, \qquad r = \xi ~, \qquad z = Z(\xi) ~ , \qquad \Omega^p = \psi^p ~,
\end{equation}
with $\psi^p$ coordinates in the $(D-3)$ unit sphere, $\xi \in (0, R)$, and $Z(R) \to 0$. Basis vectors tangent to the surface are then:
\begin{equation} \label{GeometryDisk:AdSTangentBasis}
m_1 = m_\xi = \partial_r + Z' \partial_z ~, \qquad m_{p+1} = m_{\psi^p} = \partial_{\Omega^p} ~ .
\end{equation}
This induces a metric on the surface of the form:
\begin{equation} \label{GeometryDisk:AdSTangentMetric}
\mathrm{d}s_{\mathcal{C}_1}^2 = \frac{L_{\star}^2}{Z^2(\xi)} \left[ \left( 1 + Z'^2 \right) \mathrm{d}\xi^2 + \xi^2 \, \mathrm{d}\Omega^2_{D-3} \right] ~ .
\end{equation}
We then choose our two normalized vectors orthogonal to the surface to be:
\begin{equation} \label{GeometryDisk:AdSNormalBasis}
n_1 = \frac{Z}{L_{\star}} \partial_{\tau} ~, \quad n_2 = \frac{Z}{L_{\star} \sqrt{1 + Z'^2}} \left( Z' \partial_r - \partial_z \right) ~ .
\end{equation}
This produces the following extrinsic curvature components:
\begin{align}
K^2{}_{\xi \xi} & = \frac{L_{\star}}{Z^2 \sqrt{1 + Z'^2}}\left[ 1 + Z'^2 + Z Z'' \right] ~ , \\
K^2{}_{\psi^p \psi^q} & = \frac{L_{\star}}{Z^2 \sqrt{1 + Z'^2}}  \left( \xi + Z Z' \right) \xi \, \gamma_{p q} ~ ,
\end{align}
where $\gamma_{p q}$ is the metric on the unit round sphere and the rest of the components of the extrinsic curvature vanish. Transforming the last two indices to coordinate ones:
\begin{equation*}
K^2{}_{zz} = \frac{Z'^2}{(1 + Z'^2)^2} K^2{}_{\xi \xi} ~, \quad K^2{}_{zr} = \frac{Z'}{(1 + Z'^2)^2} K^2{}_{\xi \xi} ~, \quad K^2{}_{rr} = \frac{1}{(1 + Z'^2)^2} K^2{}_{\xi \xi} ~ , 
\end{equation*}
while $K^2{}_{pq} = K^2{}_{\psi^p \psi^q}$. The trace of the second extrinsic curvature is then (clearly $K^1 = 0$):
\begin{equation}
K^2 = \frac{1}{L_{\star} \xi (1 + Z'^2)^{3/2}} \left[ \xi Z Z'' + (D-3) Z Z' \left( 1 + Z'^2 \right) + (D-2) \xi \left( 1 + Z'^2 \right) \right] ~ .
\end{equation}
Notice that the RT surface satisfying the prescribed boundary conditions (which is the one solving $K^2 = 0$) is given by $Z(\xi) = \sqrt{R^2 - \xi^2}$. In particular, this implies $Z Z' = - \xi$ and $Z Z'' = - (1 + Z'^2)$, showing that not only $K^2 = 0$, but the whole tensor satisfies $K^2{}_{\mu \nu} = 0$. This fact is relevant when discussing minimal surfaces for boundary disks in the main body of the paper.

\section{Geometry of the entanglement surface for a boundary strip}
\label{app:GeometryStrip}

The aim of this short appendix is to collect the results needed to evaluate the tensors appearing in the entanglement entropy functional for a boundary strip in Poincar\'e AdS. Consider then the bulk metric:
\begin{equation} \label{GeometryStrip:AdSMetric}
\mathrm{d}s^2 = \frac{L_{\star}^2}{z^2} \left( \mathrm{d}\tau^2 + \mathrm{d}z^2 + \mathrm{d}x^2 + \delta_{p q} \, \mathrm{d}y^p \mathrm{d} y^q \right) ~ ,
\end{equation}
where the length scale $L_{\star}$ is determined by imposing this to be a solution of the equations of motion for the theory at hand. We have separated the spatial coordinates in the boundary into $x$ and $y^p$ (with $p = 1, 2 \dots, D-3$) because we will consider in this spacetime a surface anchored to a boundary strip finite in extent in the $x$-direction, parametrized as:
\begin{equation} \label{GeometryStrip:AdSStrip}
\tau = \tau_0 ~, \qquad z = Z(\xi) ~ , \qquad x = X(\xi) ~, \qquad y^p = \psi^p ~,
\end{equation}
with $\psi^p \in (- \infty, \infty)$, $\xi \in (\xi_i, \xi_f)$, $X(\xi_i) \to - \ell/2$, $X(\xi_f) \to \ell/2$, and $Z(\xi_i), Z(\xi_f) \to 0$. Basis vectors tangent to the surface are then:
\begin{equation} \label{TensorsSolutions:AdSTangentBasis}
m_1 = m_\xi = X' \partial_x + Z' \partial_z ~, \qquad m_{p+1} = m_{\psi^p} = \partial_{y^p} ~ .
\end{equation}
This induces a metric on the surface of the form:
\begin{equation} \label{TensorsSolutions:AdSTangentMetric}
\mathrm{d}s_{\mathcal{C}_1}^2 = \frac{L_{\star}^2}{Z^2(\xi)} \left[ \left( X'^2 + Z'^2 \right) \mathrm{d}\xi^2 + \delta_{p q} \, \mathrm{d}\psi^p \mathrm{d}\psi^q \right] ~ .
\end{equation}
We then choose our two normalized vectors orthogonal to the surface to be:
\begin{equation} \label{TensorsSolutions:AdSNormalBasis}
n_1 = \frac{Z}{L_{\star}} \partial_{\tau} ~, \quad n_2 = \frac{Z}{L_{\star} \sqrt{X'^2 + Z'^2}} \left( Z' \partial_x - X' \partial_z \right) ~ .
\end{equation}
This produces the following extrinsic curvature components:
\begin{align}
K^2{}_{\xi \xi} & = \frac{L_{\star} X'^2}{Z^2 \sqrt{X'^2 + Z'^2}}\left[ X' + \left( \frac{Z Z'}{X'} \right)' \right] ~ , \\
K^2{}_{\psi^p \psi^q} & = \frac{L_{\star} X'}{Z^2 \sqrt{X'^2 + Z'^2}} \, \delta_{p q} ~ ,
\end{align}
with the rest of them vanishing. Transforming the last two indices to coordinate ones:
\begin{equation*}
K^2{}_{zz} = \frac{Z'^2}{(X'^2 + Z'^2)^2} K^2{}_{\xi \xi} ~, \quad K^2{}_{zx} = \frac{X' Z'}{(X'^2 + Z'^2)^2} K^2{}_{\xi \xi} ~, \quad K^2{}_{xx} = \frac{X'^2}{(X'^2 + Z'^2)^2} K^2{}_{\xi \xi} ~ , 
\end{equation*}
while $K^2{}_{pq} = K^2{}_{\psi^p \psi^q}$. The trace of the second extrinsic curvature is then (clearly $K^1 = 0$):
\begin{equation}
K^2 = \frac{X'}{L_{\star} (X'^2 + Z'^2)^{3/2}} \left[ (D-2) \left( X'^2 + Z'^2 \right) + Z X' \left( \frac{Z'}{X'} \right)' \right] ~ .
\end{equation}
As a final comment, we are employing a generic parametrization, but the results in the main text are presented with $z = Z(x)$. For that case, we can just set $X' = 1$. For numerical computations we also employed $x = X(z)$, in which case we set $Z = z$ and $Z' = 1$.

\section{Details of the  HEE in  ECG calculation}
\label{app:MinimizationFunctionals}

This appendix contains the calculational details of the example presented in section \ref{sec:StripECG}. First we will obtain the equations to solve, and then present details of the numerical integration. 

\subsection{Minimizing the entropy functional for a strip in ECG}
Consider the boundary strip in a state dual to the 4-dimensional vacuum AdS in ECG, with bulk metric given by \eqref{StripECG:AdS_metric}. For this setup, the functionals \eqref{FunctionalCubic:functional_XD} and \eqref{FunctionalCubic:functional_JC} have been already obtained in the minimal and non-minimal prescriptions for the $z=Z(x)$ parametrization, eqs. \eqref{StripECG:functional_minimal}--\eqref{StripECG:S_non-minimal}. Similar expressions are needed for the $x=X(z)$ parametrization, since it will be used in the first two parts of the numerical procedure. With the minimal splitting regularization we obtain
\begin{equation}
S^{min}_{strip} = \frac{L_{\star}^2}{4 G_N} \int \mathrm{d} y \, \mathrm{d} z \left[ \frac{\sqrt{1 + X'^2}}{z^2} + \frac{3 f_{\infty}^2 \mu}{4 z^2 \left[ 1 + X'^2 \right]^{9/2}} \; \mathcal{S}^{min}_x \right] ~ , 
\end{equation}
where
\begin{align} \label{MinimizationFunctionals:XS_minimal}
\mathcal{S}^{min}_x \equiv &~ 4-14 X'^6 - 18 X'^8 - 6 X'^{10} + 28 z X'^3 X'' + 32 z X'^5 X'' + 12 z X'^7 X''\nonumber \\
 & +X'^4 \left(6-3 z^2 X''^2 \right)-3 X'^2 \left(-4+z^2 X''^2 \right) + z X' X'' \left(8+z^2 X''^2 \right)~.
\end{align}
Minimizing this functional we obtain a fourth order differential equation for $X(z)$ to solve, 
\begin{align} \label{MinimizationFunctionals:XDeqX}
&4 \left[ 2 X' + 2 (X')^3 - z X'' \right] \left[ 1 + (X')^2 \right]^5 - 3 \mu f_{\infty}^2 \left[ 176 (X')^9 + 72 (X')^{11}-6 z (X')^{10} X'' \right. \nonumber \\
&+12 (X')^{13}+ (X')^7 \left(224-60 z^3 X'' X^{(3)} \right) + z X'' \left(-4+z^2 (X'')^2-12 z^3 X'' X^{(3)} \right)  \nonumber \\
&+ (X')^8 \left(-38 z X'' + 6 z^3 X^{(4)} \right) + 2 (X')^6 \left(-41 z X'' + 45 z^3 (X'')^3 + 9 z^3 X^{(4)} \right) + \nonumber \\
& + (X')^4 \left(-78 z X'' + 37 z^3 (X'')^3 + 96 z^4 (X'')^3 X^{(3)} + 18 z^3 X^{(4)} \right) + \nonumber \\
&+ (X')^2 \left(-32 z X'' - 52 z^3 (X'')^3 + 84 z^4 (X'')^2 X^{(3)} + 6 z^3 X^{(4)} \right) + \nonumber \\
& + 2 X' \left(4 + 27 z^4 (X'')^4 - 3 z^4 (X^{(3)})^2 - 3 z^3 X'' \left[ -2 X^{(3)} + z X^{(4)} \right] \right) \nonumber \\
& - 4 (X')^3 \left( -14 + 36 z^4 (X'')^4 + 3 z^4 (X^{(3)})^2 + 3 z^3 X'' \left[ 3 X^{(3)} + z X^{(4)} \right] \right) \nonumber \\
& \left. - 6 (X')^5 \left(-26 + z^4 (X^{(3)})^2 + z^3 X'' \left[18 X^{(3)} + z X^{(4)} \right] \right)  \right] = 0 ~.
\end{align}
If we use the non-minimal splitting instead, the functional obtained is
\begin{equation} \label{MinimizationFunctionals:Xfunctional_non-minimal}
S^{non-min}_{strip} = \frac{L_{\star}^2}{4 G_N} \int \mathrm{d} y \, \mathrm{d} z \left[ \frac{\sqrt{1 + X'^2}}{z^2} + \frac{3 f_{\infty}^2 \mu}{8 z^2 \left[ 1 + X'^2 \right]^{9/2}} \; \mathcal{S}^{non-min}_x \right] ~.
\end{equation}
where
\begin{align} \label{MinimizationFunctionals:XS_non-minimal}
\mathcal{S}^{non-min}_x \equiv  &~ 8 -10 X'^6 -18   X'^8 - 6 X'^{10} +44 z X'^3 X'' + 40 z X'^5 X'' + 12 z X'^7 X'' \nonumber \\
& +z X' X'' \left( 16+3 z^2 X''^2 \right) + X'^2 \left( 24 - z^2 X''^2\right) + X'^4 \left( 18- z^2 X''^2 \right) .
\end{align}
Minimizing this functional, the equation to solve is
\begin{align} \label{MinimizationFunctionals:RXMeqX}
&8 \left[ 2 X' + 2 (X')^3 - z X'' \right] \left[ 1 + (X')^2 \right]^5 + 3 \mu f_{\infty}^2 \left[ -184 (X')^9 - 72 (X')^{11} + 6 z (X')^{10} X'' \right. \nonumber \\
&-12 (X')^{13}+ 4 (X')^7 \left(-64 + 5 z^3 X'' X^{(3)} \right) + z X'' \left(8+13 z^2 (X'')^2 + 36 z^3 X'' X^{(3)} \right)+  \nonumber \\
&+ (X')^8 \left(22 z X'' - 2  z^3 X^{(4)} \right) + 2 (X')^6 \left(38 z X'' -30 z^3 (X'')^3 -6 z^3 X^{(4)} \right) - \nonumber \\
& - (X')^4 \left(-42 z X'' + 119 z^3 (X'')^3 + 228 z^4 (X'')^2 X^{(3)} + 6 z^3 X^{(4)} \right) - \nonumber \\
&-2 (X')^2 \left(-14 z X'' + 38 z^3 (X'')^3 + 126 z^4 (X'')^2 X^{(3)} + z^3 X^{(4)} \right) - \nonumber \\
& - 2 X' \left(8 + 81 z^4 (X'')^4 - 9 z^4 (X^{(3)})^2 - z^3 X'' \left[ 14 X^{(3)} + 9 z X^{(4)} \right] \right) + \nonumber \\
& + 4 (X')^3 \left( -22 + 108 z^4 (X'')^4 + 9 z^4 (X^{(3)})^2 + z^3 X'' \left[ 19 X^{(3)} + 9 z X^{(4)} \right] \right)+ \nonumber \\
& \left. +26 (X')^5 \left(-102 + 9 z^4 (X^{(3)})^2 + z^3 X'' \left[34 X^{(3)} + 9 z X^{(4)} \right] \right)  \right] = 0 ~.
\end{align}
It will prove convenient for the numerical calculation to work also with the entanglement functional in terms of $Z(x)$ instead of $X(z)$. Recall that, with the minimal prescription, the functional reads:
\begin{equation} \label{MinimizationFunctionals:functional_minimal}
S^{min}_{strip} = \frac{L_{\star}^2}{4 G_N} \int \mathrm{d} y \, \mathrm{d} x \left[ \frac{\sqrt{1 + (Z')^2}}{Z^2} + \frac{3 f_{\infty}^2 \mu}{4 Z^2 \left[ 1 + (Z')^2 \right]^{9/2}} \; \mathcal{S}^{min} \right] ~ , 
\end{equation}
where
\begin{align} \label{MinimizationFunctionals:S_minimal}
\mathcal{S}^{min} \equiv & - 6 + 12 (Z')^8 + 4 (Z')^{10} - 12 Z Z'' - 3 Z^2 (Z'')^2 - Z^3 (Z'')^3 + (Z')^6 \left( 6 - 8 Z Z'' \right) - \nonumber \\
 & - 14 (Z')^4 \left( 1 + 2 Z Z'' \right) - (Z')^2 \left[ 18 + 32 Z Z'' + 3 Z^2 (Z'')^2 \right] ~.
\end{align}
And following the non-minimal prescription:
\begin{equation} \label{MinimizationFunctionals:functional_non-minimal}
S^{non-min}_{strip} = \frac{L_{\star}^2}{4 G_N} \int \mathrm{d} y \, \mathrm{d} x \left[ \frac{\sqrt{1 + (Z')^2}}{Z^2} + \frac{3 f_{\infty}^2 \mu}{8 Z^2 \left[ 1 + (Z')^2 \right]^{9/2}} \; \mathcal{S}^{non-min} \right] ~ ,
\end{equation}
where now
\begin{align} \label{MinimizationFuncitonals:S_non-minimal}
\mathcal{S}^{non-min} \equiv & -6 + 12 (Z')^8 + 4 (Z')^{10} - 12 Z Z'' - 3 Z^2 (Z'')^2 -  Z^3 (Z'')^3 + (Z')^6 \left( 6 - 8 Z Z'' \right) - \nonumber \\
 & - 14 (Z')^4 \left( 1 + 2 Z Z'' \right) - (Z')^2 \left( 18 + 32 Z Z'' + 3 Z^2 (Z'')^2 \right)  ~.
\end{align}
Then, the equations to solve are 
\begin{align} \label{MinimizationFunctionals:XDeqZ}
&4 \left[ 2 + 2 (Z')^2 + Z Z'' \right] \left[ 1 + (Z')^2 \right]^5 - 3 \mu f_{\infty}^2 \left[ 12 + 8 (Z')^{12} + 6 Z Z'' + 72 Z^3 (Z')^5 Z'' Z^{(3)} +  \right. \nonumber \\
&+ 36 Z^3 Z' Z'' Z^{(3)} \left( 2 + 3 Z Z'' \right)+ 4 (Z')^{10} \left(14 + Z Z'' \right) + 36 Z^3 (Z')^3  Z'' Z^{(3)} \left(4 + 3 Z Z'' \right)  + \nonumber \\
&+ 4 (Z')^8 \left(39 + Z Z'' \right) + Z^3 \left( 17 (Z'')^3 - 6 Z^{(4)} \right) + 6 Z^4 \left(3 (Z'')^4 - (Z^{(3)})^2 - Z'' Z^{(4)} \right) + \nonumber \\
&  + (Z')^6 \left(224 + 78 Z Z'' - 6 Z^3 Z^{(4)} \right) -(Z')^4 \left(-176 - 82 Z Z''  + \right. \nonumber \\
&  \left. + Z^3 \left[ 127 (Z'')^3 + 18 Z^{(4)} \right] + 6 Z^4 \left[ (Z^{(3)})^2 + Z'' Z^{(4)} \right] \right) -2 (Z')^2 \left( -36 - 19 Z Z'' + \right. \nonumber \\
&\left. \left. + Z^3 \left[ 55 (Z'')^3 + 9 Z^{(4)} \right] +  6 Z^4 \left[ 15 (Z'')^4 + (Z^{(3)})^2 + Z'' Z^{(4)} \right] \right)  \right] = 0 ~,
\end{align}
if we use the minimal splitting, and 
\begin{align} \label{MinimizationFunctionals:RXMeqZ}
&8 \left[ 2 + 2 (Z')^2 + Z Z'' \right] \left[ 1 + (Z')^2 \right]^5 - 3 \mu f_{\infty}^2 \left[ 12 + 16 (Z')^{12} + 6 Z Z'' + 8 Z^3 (Z')^5 Z'' Z^{(3)} +  \right. \nonumber \\
&+ 4 Z^3 Z' Z'' Z^{(3)} \left( -2 + 81 Z Z'' \right)+ 8 (Z')^{10} \left(11 + Z Z'' \right) + 4 Z^3 (Z')^3  Z'' Z^{(3)} \left(-4 + 81 Z Z'' \right)  +  \nonumber \\
&+ 4 (Z')^8 \left(51 + 7 Z Z'' \right) - Z^3 \left( 5 (Z'')^3 + 2 Z^{(4)} \right) + 18 Z^4 \left(3 (Z'')^4 - (Z^{(3)})^2 - Z'' Z^{(4)} \right) + \nonumber \\
&  + (Z')^6 \left(256 + 42 Z Z'' - 2 Z^3 Z^{(4)} \right) + (Z')^4 \left(184 + 28 Z Z''  + \right. \nonumber \\
&  \left. + Z^3 \left[ 67 (Z'')^3 - 6 Z^{(4)} \right] -18  Z^4 \left[ (Z^{(3)})^2 + Z'' Z^{(4)} \right] \right) + (Z')^2 \left( 71 +22 Z Z'' + \right. \nonumber \\
&\left. \left. + Z^3 \left[ 62 (Z'')^3 - 6 Z^{(4)} \right] - 36 Z^4 \left[ 15 (Z'')^4 + (Z^{(3)})^2 + Z'' Z^{(4)} \right] \right)  \right] = 0 .
\end{align}
when using the non-minimal splitting. 
%

\subsection{Numerics}
\label{app:DetailsNumerical}

We consider an  interval of width $\ell/2= 1.5$. The  $Z(x)$ parametrization is problematic if we want to start integrating from the boundary keeping the endpoints of the interval fixed. It is more convenient to work --at least initially-- in terms of $X(z)$. Our strategy will be to start with a series expansion for $X(z)$ close to the boundary, at $X(0)=\ell/2$, then numerically integrate equation \eqref{MinimizationFunctionals:XDeqX} (or \eqref{MinimizationFunctionals:RXMeqX} depending on what splitting are we considering) and, finally, when the $X(z)$ parametrization becomes problematic, switch to $Z(x)$ and numerically integrate \eqref{MinimizationFunctionals:XDeqZ} (or \eqref{MinimizationFunctionals:RXMeqZ}) until $x=0$, where we reach the deepest point of the curve, $Z(0)=z_*$.

\subsubsection{Series expansion close to the boundary}

\begin{figure}[!ht]

\begin{subfigure}{0.5\textwidth}
\includegraphics[width=1\linewidth, height=4.5cm]{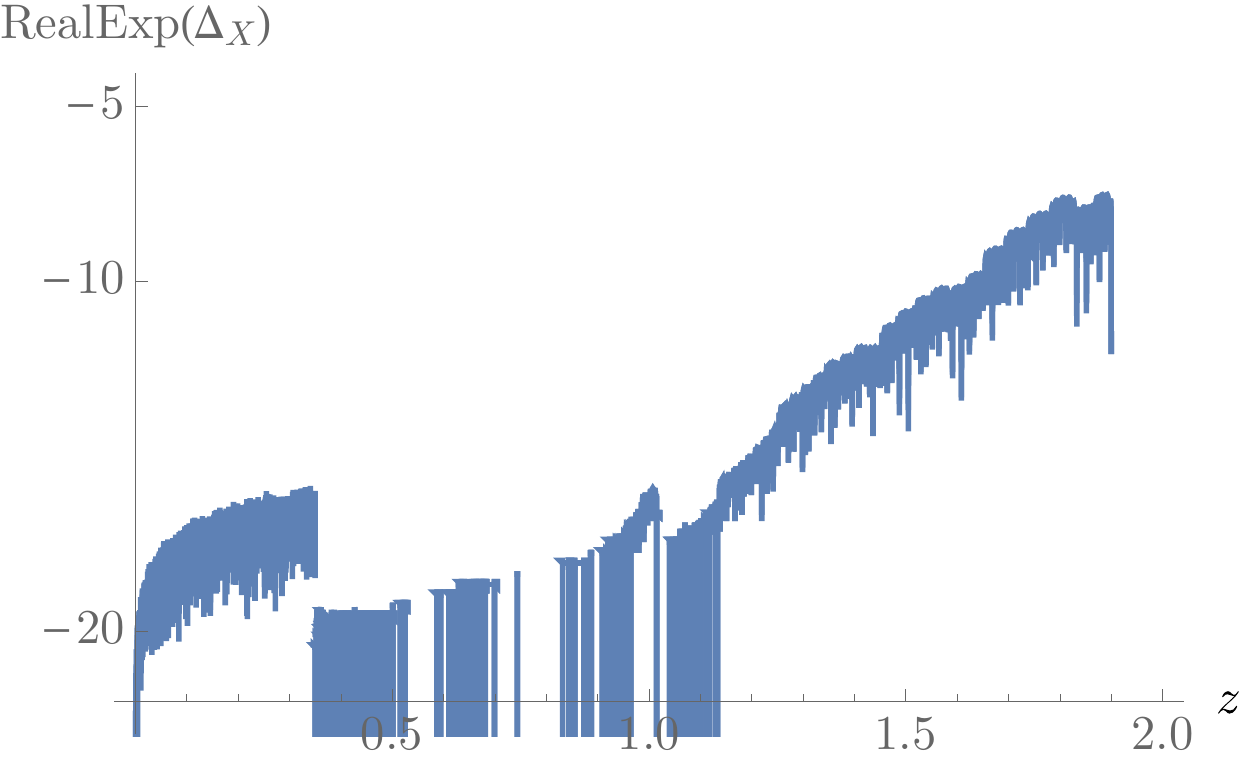} 
\caption{\footnotesize{Series expansion of $X(z)$ for $0 \leq z \leq 0.35$.    Numerical integration of $X(z)$ for $0.35 \leq z \leq 1.90$.}}
\label{FigNR:Error1-0_002}
\end{subfigure}
\begin{subfigure}{0.5\textwidth}
\includegraphics[width=1\linewidth, height=4.5cm]{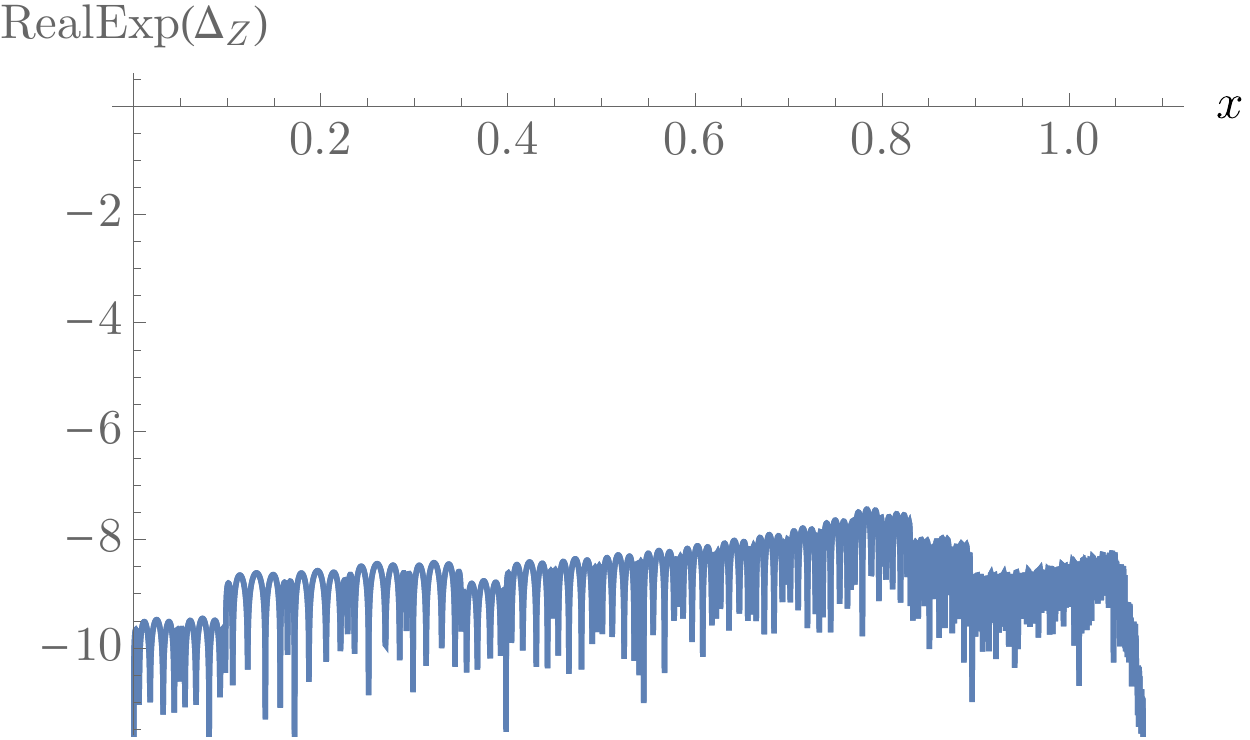}
\caption{\footnotesize{Numerical integration of $Z(x)$, $0 \leq x \leq 1.087$}}
\label{FigNR:Error3-0_002}
\end{subfigure}

\caption{\footnotesize{Order of magnitude of $\Delta$, the result obtained by evaluating the numerical solution into the corresponding differential equation, for $\mu = -0.002$ in the non-minimal case. Similar plots were obtained for all the other values of $\mu$ that were considered, for both prescriptions.}}
\label{FigNR:ErrorJC-0_002}
\end{figure}

To solve the relevant equation (\eqref{MinimizationFunctionals:XDeqX} or \eqref{MinimizationFunctionals:RXMeqX}) starting from the boundary, we perform a series expansion of $X(z)$ and the corresponding differential equation to order 23 for values close to $z=0$. The zeroth-order term is determined by the boundary condition  $X(z=0) = \ell/2$, whereas all the other coefficients depend on the value of the third-order coefficient in the expansion, related to $X'''(0)$ as well as the value of $\mu$. This series expansion is a good solution up to some small  $z=\epsilon$. The value of $\epsilon$  is chosen such that the numerical error is less than  $10^{-5}$ at any point $ 0\le z \le\epsilon$, see Figure \ref{FigNR:Error1-0_002} for a representative case.

 \subsubsection{Numerical integration}

For $\epsilon \le z \le z_I$, we integrated numerically the differential equation for $X(z)$. 
The value $z_I$ was determined for each case as the largest value of $z$ for which the errors remained below  $10^{-5}$ at any point in the interval of integration -- see figure \ref{FigNR:Error3-0_002}. At the point $(x_I,z_I)$, we changed parametrization to $z=Z(x)$ and we integrated  numerically until we reached the  $z$ axis, thus completing the solution.

\begin{figure}[h]
\centering
	\includegraphics[scale=0.41]{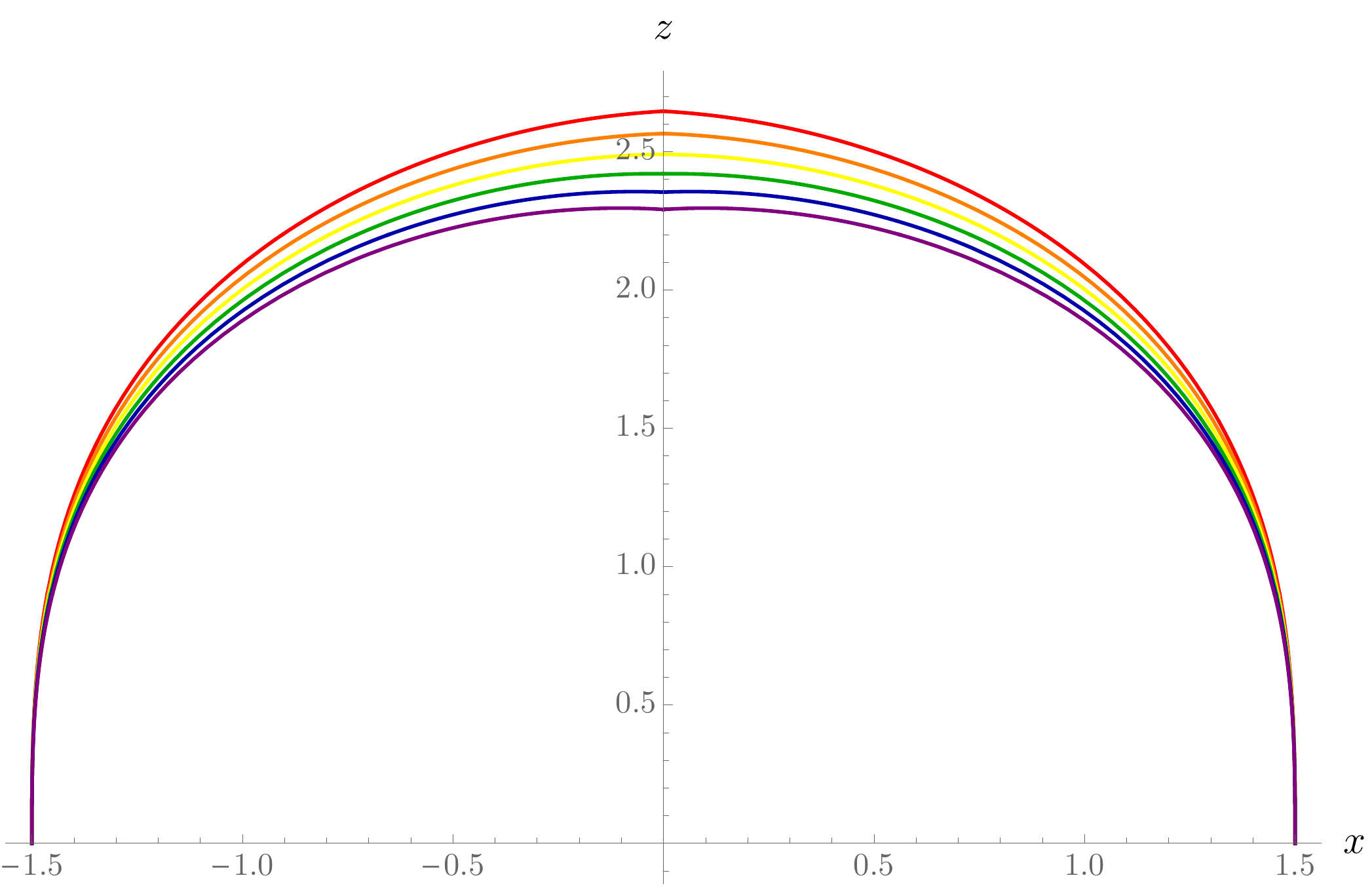}
	\caption{A family of curves anchored at $x=\pm \ell/2$ for $\mu = -0.003$ in the non-minimal prescription. This set of curves was obtained by varying the initial parameter (related to the third derivative $X'''(z)$ at $z=0$) from -0.30 to -0.40 in steps of -0.02. The curve identified as the actual entanglement surface for this case has an initial parameter close to -0.346.}
	\label{FigNR:FamilyMu}
\end{figure}

Recall that the value of the third derivative in the series expansion at $z=0$ is so far an initial free parameter, so for a single value of $\mu$ we obtain a family of solutions characterized by different values of the third derivative. In figure \ref{FigNR:FamilyMu}, each curve corresponds to a different choice for the initial parameter. Note that this added complication arises because we are dealing with fourth order differential equations, unlike the second order ones we encounter in Einstein and Gauss-Bonnet gravity. All these solutions are good solutions of the differential equation. However, in a holographic context we expect that the curve will be smooth at $x=0$. Therefore, among all the possible values of the third derivative we have to choose the one that produces a curve with $Z'(x=0)=0$, and this curve will be the RT surface. In Table \ref{TableNR:zstar} we list the deepest point of the RT surface, $z_*$, obtained for different values of the ECG coupling $\mu$. 
\begin{table}[h]
	\center
	\begin{tabular}{|c|c|c|}
		\hline 
		$\mu$ & $z_*$ Minimal prescription & $z_*$ Non-minimal prescription \\ 
		\hline
		$-10^4$  & 2.05742 & 1.84623 \\
		\hline
		$-10$  & 2.08559  & 1.87208 \\
		\hline
		$-0.50$  & 2.16944  & 1.95529  \\
		\hline
		$-0.003$ &2.50557 & 2.46914 \\ 
		\hline 
		$-0.002$ & 2.50837  & 2.48912 \\ 
		\hline 
		$-0.001$ & 2.50798 & 2.50665 \\ 
		\hline 
		$+0.003$ & 2.42368 & 2.7211 \\ 
		\hline 
		$+0.010$ & 2.64287 & 2.75594 \\
		\hline
	\end{tabular}
	\caption{Values of $z_*$ obtained from numerical solutions, for different values of $\mu$.} 
	\label{TableNR:zstar}
\end{table}

\begin{figure}[!h]
\begin{subfigure}{0.5\textwidth}
\includegraphics[width=0.9\linewidth, height=5cm]{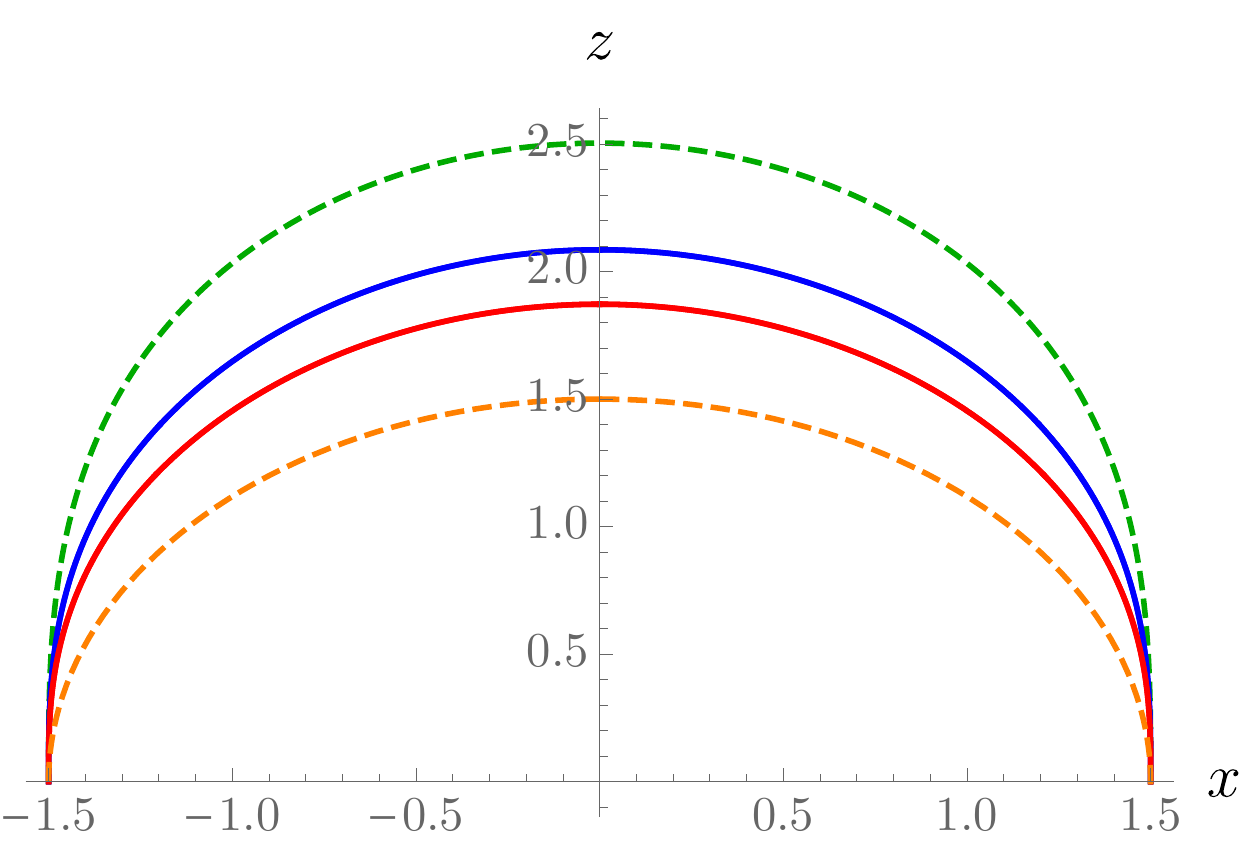}
\caption{$\mu = -10$}
\label{FigNR:Mu-10}
\end{subfigure}
\begin{subfigure}{0.5\textwidth}
\includegraphics[width=0.9\linewidth, height=5cm]{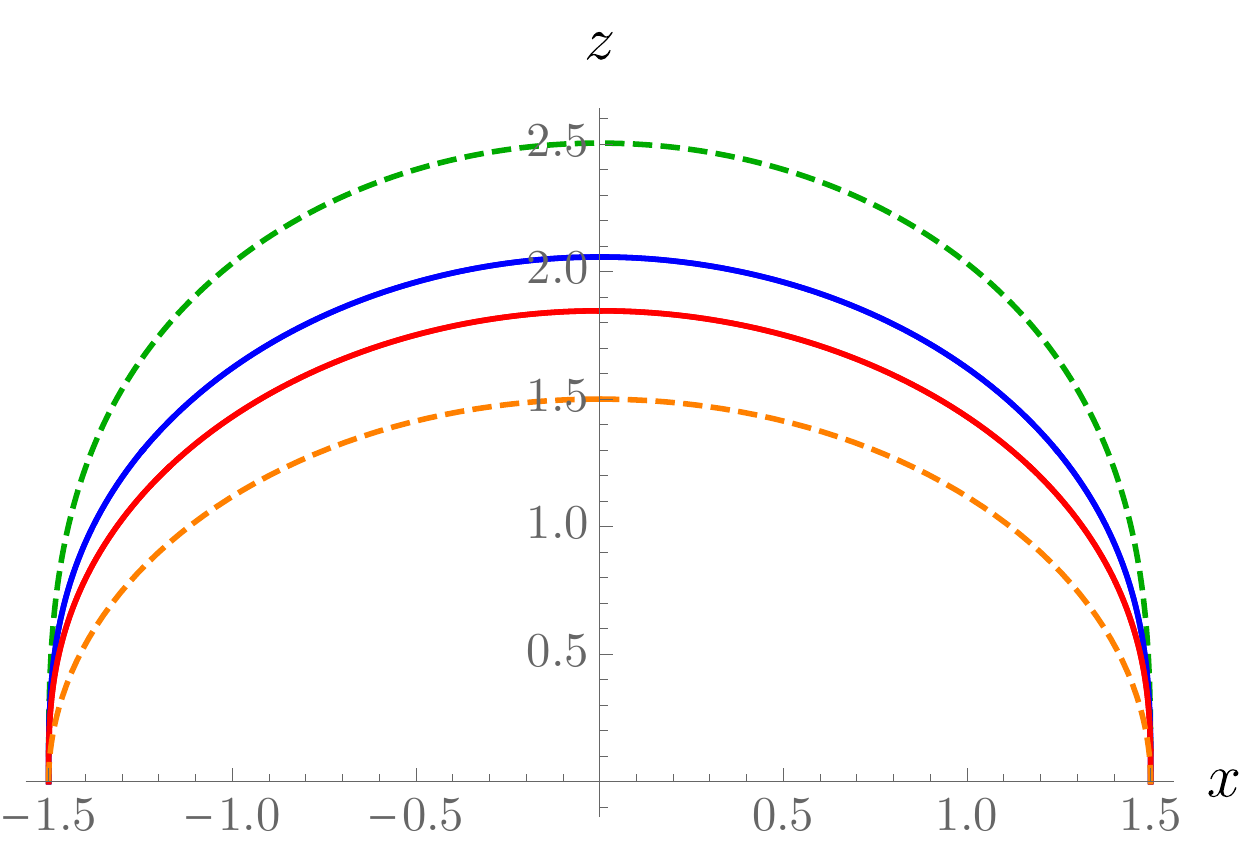}
\caption{$\mu = -10^4$}
\label{FigNR:Mu-10000}
\end{subfigure}

\caption{Causal surface (orange, dashed) and entanglement surfaces corresponding to ECG in the case of minimal prescription (blue), ECG in the case of non-minimal prescription (red), and Einstein gravity (green, dashed) \cite{HubenySurfaces} for the boundary strip length $\ell/2 =1.5$ and different values of $\mu$ outside the interval $-0.00322 \leq \mu \leq 0.00312$. In both cases, we verify that $z_c<z_*$ is satisfied for both prescriptions.}
\label{FigNR:OtherMuApp}
\end{figure}

\begin{figure}[!ht]

\begin{subfigure}{0.5\textwidth}
\includegraphics[width=0.85\linewidth, height=4.5cm]{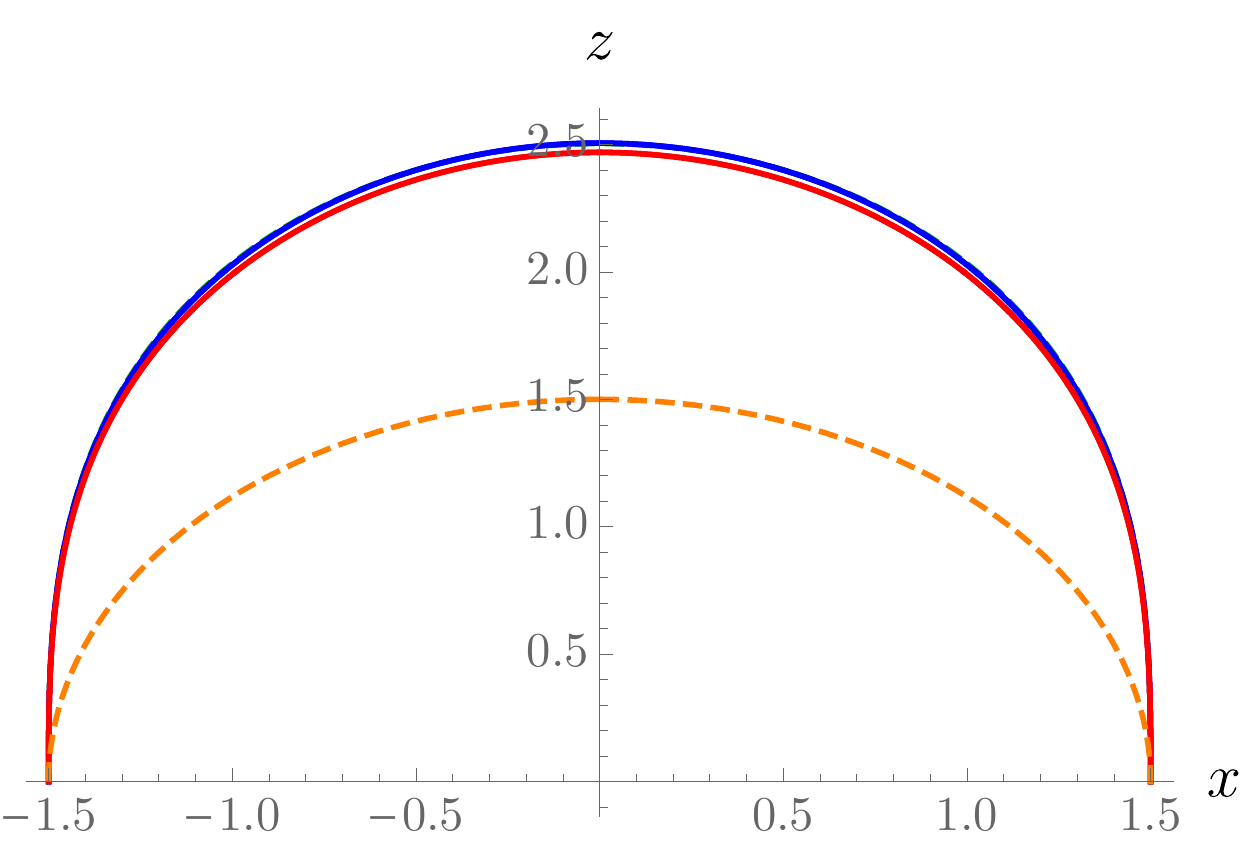} 
\caption{\footnotesize{$\mu=-0.003$}}
\label{FigNR:Mu-0_003}
\end{subfigure}
\begin{subfigure}{0.5\textwidth}
\includegraphics[width=0.85\linewidth, height=4.5cm]{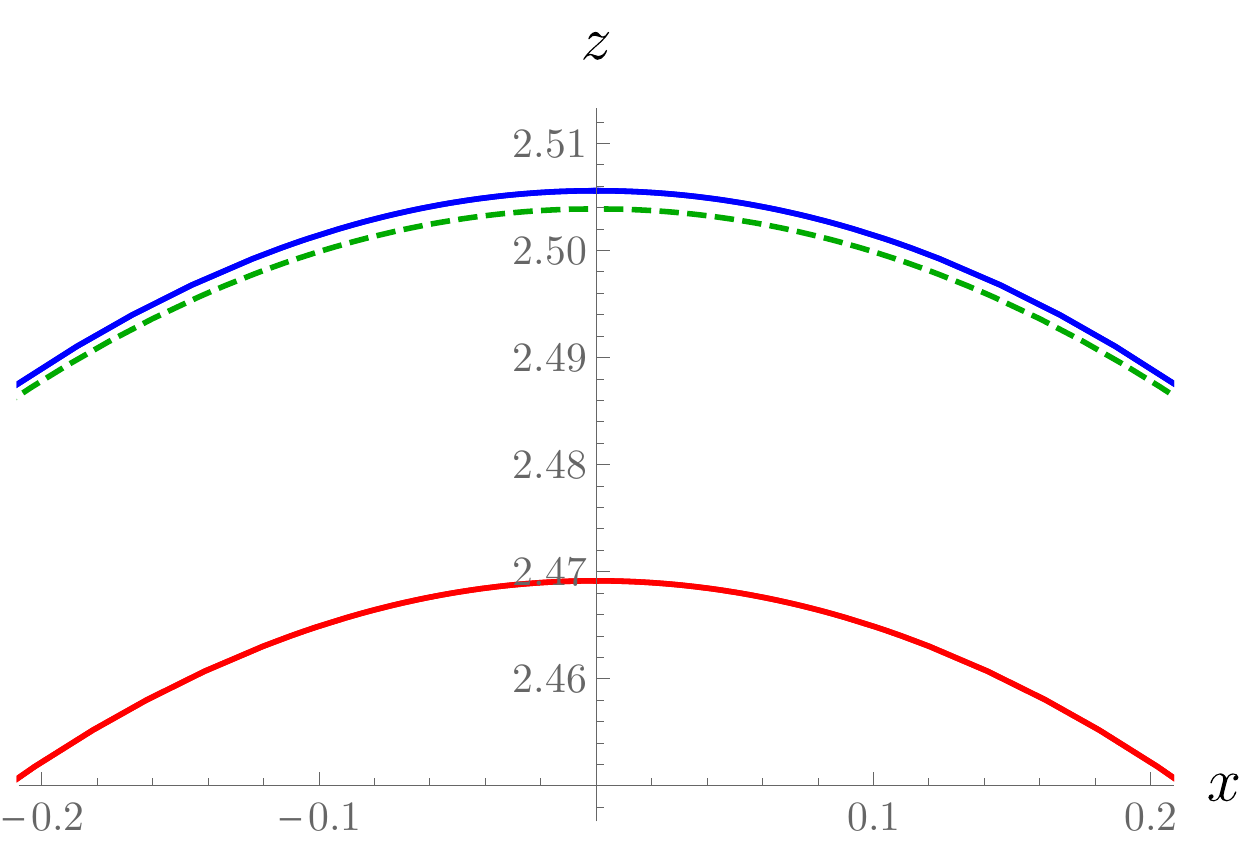}
\caption{\footnotesize{$\mu=-0.003$}}
\label{FigNR:Mu-0_003Zoomin}
\end{subfigure}

\begin{subfigure}{0.5\textwidth}
\includegraphics[width=0.85\linewidth, height=4.5cm]{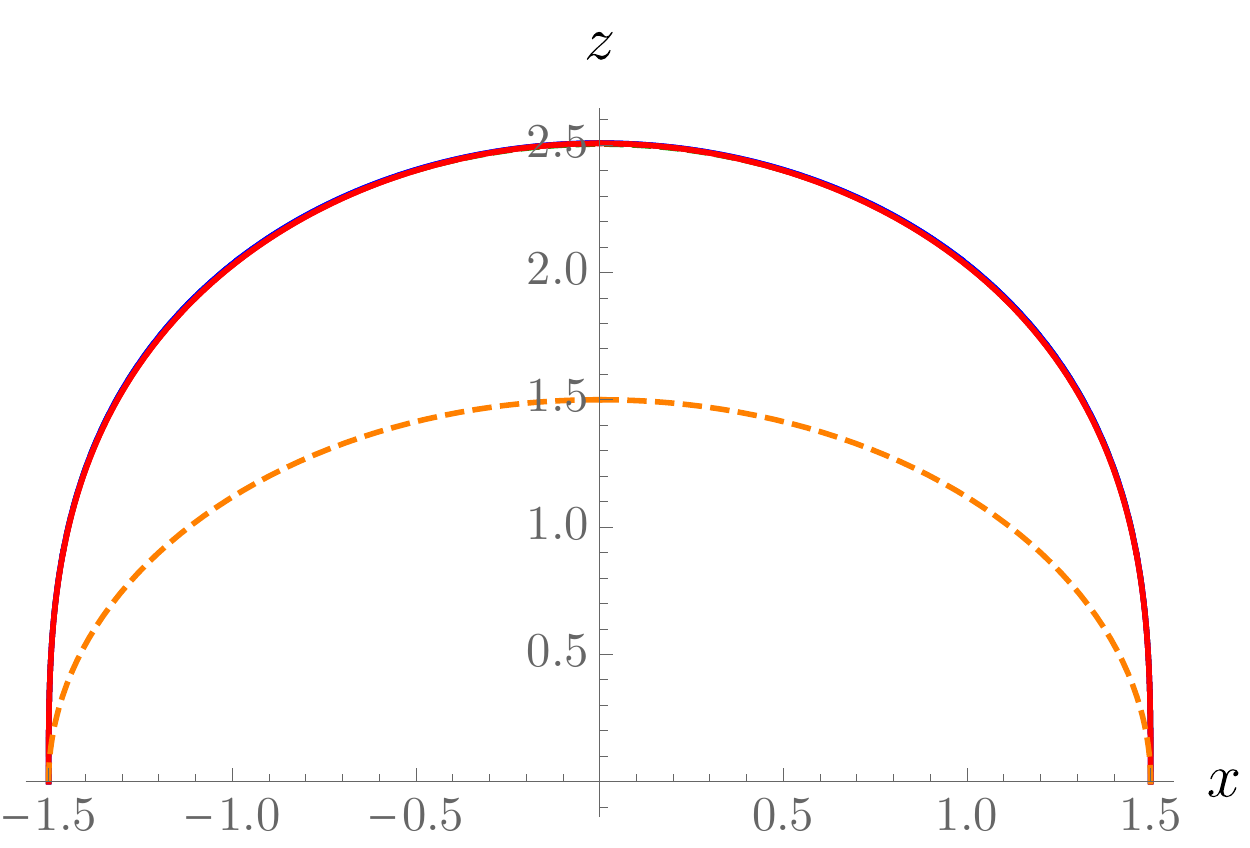} 
\caption{\footnotesize{$\mu=-0.001$}}
\label{FigNR:Mu-0_001}
\end{subfigure}
\begin{subfigure}{0.5\textwidth}
\includegraphics[width=0.85\linewidth, height=4.5cm]{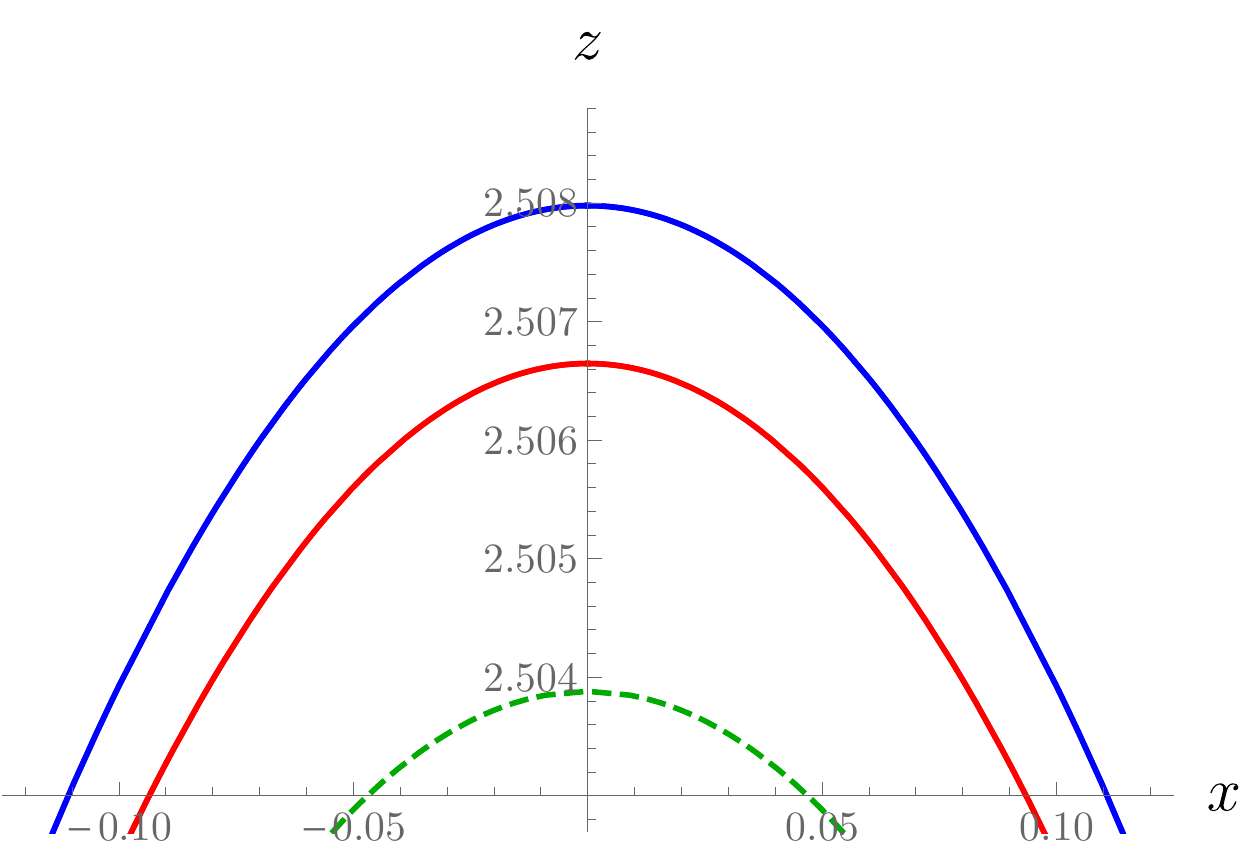}
\caption{\footnotesize{$\mu=-0.001$}}
\label{FigNR:Mu-0_001Zoomin}
\end{subfigure}

\centering
\begin{subfigure}{0.5\textwidth}
\includegraphics[width=0.9\linewidth, height=4.8cm]{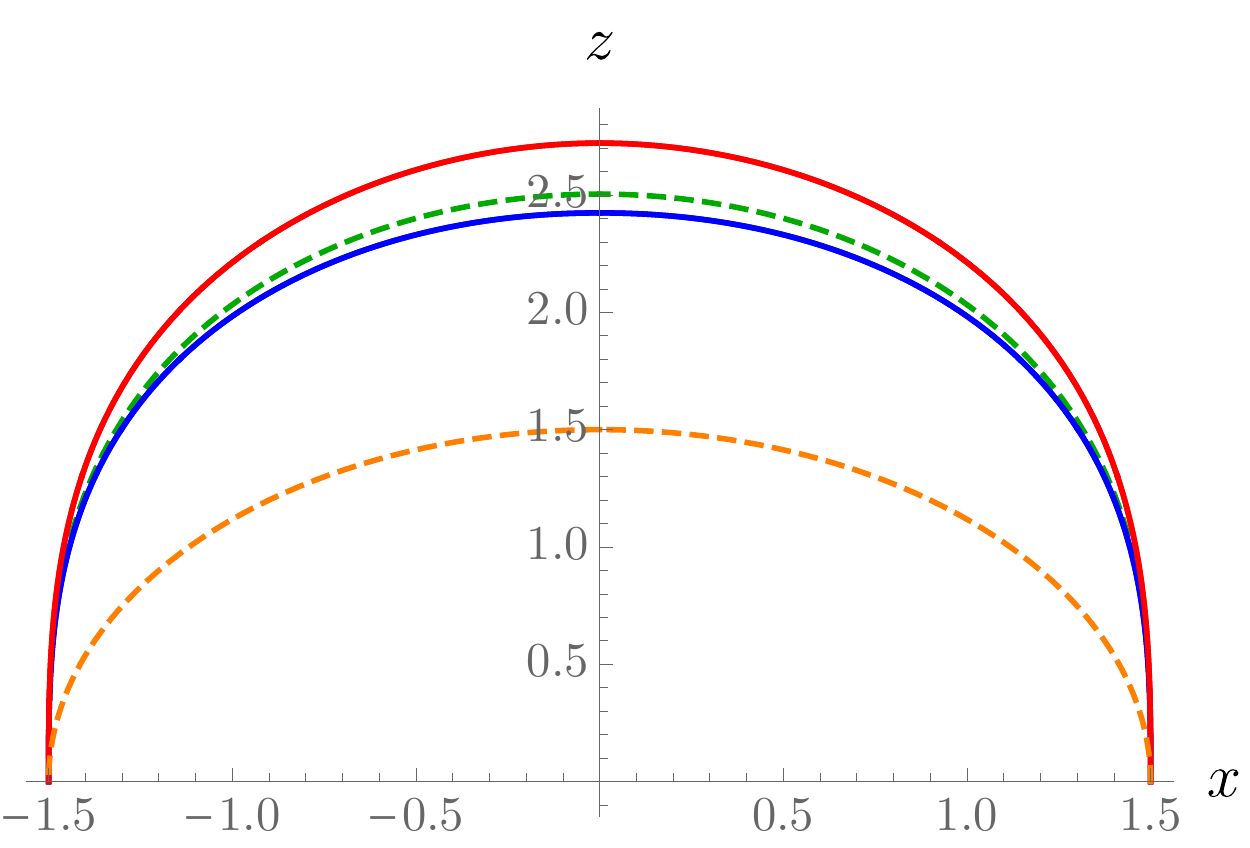}
\caption{\footnotesize{$\mu = +0.003$}}
\label{FigNR:Mu+0_003}
\end{subfigure}

\caption{\footnotesize{Causal wedge (orange, dashed) and entanglement surfaces corresponding to ECG in the case of minimal prescription (blue), ECG in the case of non-minimal prescription (red), and Einstein gravity (green, dashed) \cite{HubenySurfaces} for the boundary strip length $\ell/2 =1.5$ and different values of the coupling constant in the interval $-0.00322 \leq \mu \leq 0.00312$. For negative values of $\mu$, we include a close-up of the entanglement surfaces near $x=0$ to tell them apart better.}}
\label{FigNR:SmallerIntervalMuApp}
\end{figure}

The causal wedge in AdS is known to be a semicircle, thus, for a boundary region with $\ell/2=1.5$ the deepest point of penetration of the causal wedge is $z_c=1.5$. If, for any value of $\mu$ considered, we find that $z_*<z_c$, this would be a clear violation of the causal wedge inclusion. For all the $\mu$ values for which the solution was found there is no indication of such violation, see Table \ref{TableNR:zstar}. We also show the plots for various cases of $\mu$ values for which the solution was obtained in Figures \ref{FigNR:OtherMuApp} and \ref{FigNR:SmallerIntervalMuApp}. Note that as we make $\mu$ more negative, the value of $z_*$ decreases for both prescriptions; even for $\mu = -10^4$, which is the most negative value for which the solution was obtained, we verified that $z_*>z_c$. 

\newpage

\bibliographystyle{jhep}
\bibliography{refCG}

\providecommand{\href}[2]{#2}\begingroup\raggedright\begin{thebibliography}{10}

\bibitem{Izumi:2014loa}
K.~Izumi, \emph{{Causal Structures in Gauss-Bonnet gravity}},
  \href{https://doi.org/10.1103/PhysRevD.90.044037}{\emph{Phys. Rev. D}
  {\bfseries 90} (2014) 044037}
  [\href{https://arxiv.org/abs/1406.0677}{{\ttfamily 1406.0677}}].

\bibitem{Reall:2014pwa}
H.~Reall, N.~Tanahashi and B.~Way, \emph{{Causality and Hyperbolicity of
  Lovelock Theories}},
  \href{https://doi.org/10.1088/0264-9381/31/20/205005}{\emph{Class. Quant.
  Grav.} {\bfseries 31} (2014) 205005}
  [\href{https://arxiv.org/abs/1406.3379}{{\ttfamily 1406.3379}}].

\bibitem{Hung:2011xb}
L.-Y.~Hung, R.C.~Myers and M.~Smolkin, \emph{{On Holographic Entanglement
  Entropy and Higher Curvature Gravity}},
  \href{https://doi.org/10.1007/JHEP04(2011)025}{\emph{JHEP} {\bfseries 04}
  (2011) 025} [\href{https://arxiv.org/abs/1101.5813}{{\ttfamily 1101.5813}}].

\bibitem{Dong:entanglement_entropy}
X.~Dong, \emph{{Holographic Entanglement Entropy for General Higher Derivative
  Gravity}}, \href{https://doi.org/10.1007/JHEP01(2014)044}{\emph{JHEP}
  {\bfseries 01} (2014) 044} [\href{https://arxiv.org/abs/1310.5713}{{\ttfamily
  1310.5713}}].

\bibitem{Camps:generalized_entropy}
J.~Camps, \emph{{Generalized entropy and higher derivative Gravity}},
  \href{https://doi.org/10.1007/JHEP03(2014)070}{\emph{JHEP} {\bfseries 03}
  (2014) 070} [\href{https://arxiv.org/abs/1310.6659}{{\ttfamily 1310.6659}}].

\bibitem{CampsBoundaryCones}
J.~Camps, \emph{{Gravity duals of boundary cones}},
  \href{https://doi.org/10.1007/JHEP09(2016)139}{\emph{JHEP} {\bfseries 09}
  (2016) 139} [\href{https://arxiv.org/abs/1605.08588}{{\ttfamily
  1605.08588}}].

\bibitem{Caceres:2019pok}
E.~C\'aceres, A.S.~Misobuchi and J.F.~Pedraza, \emph{{Constraining higher order
  gravities with subregion duality}},
  \href{https://doi.org/10.1007/JHEP11(2019)175}{\emph{JHEP} {\bfseries 11}
  (2019) 175} [\href{https://arxiv.org/abs/1907.08021}{{\ttfamily
  1907.08021}}].

\bibitem{Headrick:2014cta}
M.~Headrick, V.E.~Hubeny, A.~Lawrence and M.~Rangamani, \emph{{Causality \&
  holographic entanglement entropy}},
  \href{https://doi.org/10.1007/JHEP12(2014)162}{\emph{JHEP} {\bfseries 12}
  (2014) 162} [\href{https://arxiv.org/abs/1408.6300}{{\ttfamily 1408.6300}}].

\bibitem{Akers:2016ugt}
C.~Akers, J.~Koeller, S.~Leichenauer and A.~Levine, \emph{{Geometric
  Constraints from Subregion Duality Beyond the Classical Regime}},
  \href{https://arxiv.org/abs/1610.08968}{{\ttfamily 1610.08968}}.

\bibitem{Camanho:2014apa}
X.O.~Camanho, J.D.~Edelstein, J.~Maldacena and A.~Zhiboedov, \emph{{Causality
  Constraints on Corrections to the Graviton Three-Point Coupling}},
  \href{https://doi.org/10.1007/JHEP02(2016)020}{\emph{JHEP} {\bfseries 02}
  (2016) 020} [\href{https://arxiv.org/abs/1407.5597}{{\ttfamily 1407.5597}}].

\bibitem{Brigante_2008}
M.~Brigante, H.~Liu, R.C.~Myers, S.~Shenker and S.~Yaida, \emph{{Viscosity
  Bound Violation in Higher Derivative Gravity}},
  \href{https://doi.org/10.1103/PhysRevD.77.126006}{\emph{Phys. Rev. D}
  {\bfseries 77} (2008) 126006}
  [\href{https://arxiv.org/abs/0712.0805}{{\ttfamily 0712.0805}}].

\bibitem{Buchel_2009}
A.~Buchel, R.C.~Myers and A.~Sinha, \emph{{Beyond eta/s = 1/4 pi}},
  \href{https://doi.org/10.1088/1126-6708/2009/03/084}{\emph{JHEP} {\bfseries
  03} (2009) 084} [\href{https://arxiv.org/abs/0812.2521}{{\ttfamily
  0812.2521}}].

\bibitem{Chen:2020uac}
H.Z.~Chen, R.C.~Myers, D.~Neuenfeld, I.A.~Reyes and J.~Sandor, \emph{{Quantum
  Extremal Islands Made Easy, Part I: Entanglement on the Brane}},
  \href{https://arxiv.org/abs/2006.04851}{{\ttfamily 2006.04851}}.

\bibitem{Oliva:2010eb}
J.~Oliva and S.~Ray, \emph{{A new cubic theory of gravity in five dimensions:
  Black hole, Birkhoff's theorem and C-function}},
  \href{https://doi.org/10.1088/0264-9381/27/22/225002}{\emph{Class. Quant.
  Grav.} {\bfseries 27} (2010) 225002}
  [\href{https://arxiv.org/abs/1003.4773}{{\ttfamily 1003.4773}}].

\bibitem{PablosECG}
P.~Bueno and P.A.~Cano, \emph{{Einsteinian cubic gravity}},
  \href{https://doi.org/10.1103/PhysRevD.94.104005}{\emph{Phys. Rev. D}
  {\bfseries 94} (2016) 104005}
  [\href{https://arxiv.org/abs/1607.06463}{{\ttfamily 1607.06463}}].

\bibitem{Myers:2010jv}
R.C.~Myers, M.F.~Paulos and A.~Sinha, \emph{{Holographic studies of
  quasi-topological gravity}},
  \href{https://doi.org/10.1007/JHEP08(2010)035}{\emph{JHEP} {\bfseries 08}
  (2010) 035} [\href{https://arxiv.org/abs/1004.2055}{{\ttfamily 1004.2055}}].

\bibitem{PablosHolography}
P.~Bueno, P.A.~Cano and A.~Ruip\'erez, \emph{{Holographic studies of
  Einsteinian cubic gravity}},
  \href{https://doi.org/10.1007/JHEP03(2018)150}{\emph{JHEP} {\bfseries 03}
  (2018) 150} [\href{https://arxiv.org/abs/1802.00018}{{\ttfamily
  1802.00018}}].

\bibitem{Bhattacharyya:2013gra}
A.~Bhattacharyya, M.~Sharma and A.~Sinha, \emph{{On generalized gravitational
  entropy, squashed cones and holography}},
  \href{https://doi.org/10.1007/JHEP01(2014)021}{\emph{JHEP} {\bfseries 01}
  (2014) 021} [\href{https://arxiv.org/abs/1308.5748}{{\ttfamily 1308.5748}}].

\bibitem{Ghodsi:2015gna}
A.~Ghodsi and M.~Moghadassi, \emph{{Holographic entanglement entropy from
  minimal surfaces with/without extrinsic curvature}},
  \href{https://doi.org/10.1007/JHEP02(2016)037}{\emph{JHEP} {\bfseries 02}
  (2016) 037} [\href{https://arxiv.org/abs/1508.02527}{{\ttfamily
  1508.02527}}].

\bibitem{MiaoSplitting1}
R.-X.~Miao and W.-z.~Guo, \emph{{Holographic Entanglement Entropy for the Most
  General Higher Derivative Gravity}},
  \href{https://doi.org/10.1007/JHEP08(2015)031}{\emph{JHEP} {\bfseries 08}
  (2015) 031} [\href{https://arxiv.org/abs/1411.5579}{{\ttfamily 1411.5579}}].

\bibitem{CampsKelly}
J.~Camps and W.R.~Kelly, \emph{{Generalized gravitational entropy without
  replica symmetry}},
  \href{https://doi.org/10.1007/JHEP03(2015)061}{\emph{JHEP} {\bfseries 03}
  (2015) 061} [\href{https://arxiv.org/abs/1412.4093}{{\ttfamily 1412.4093}}].

\bibitem{MiaoSplitting2}
R.-X.~Miao, \emph{{Universal Terms of Entanglement Entropy for 6d CFTs}},
  \href{https://doi.org/10.1007/JHEP10(2015)049}{\emph{JHEP} {\bfseries 10}
  (2015) 049} [\href{https://arxiv.org/abs/1503.05538}{{\ttfamily
  1503.05538}}].

\bibitem{LewkowyczMaldacena}
A.~Lewkowycz and J.~Maldacena, \emph{{Generalized gravitational entropy}},
  \href{https://doi.org/10.1007/JHEP08(2013)090}{\emph{JHEP} {\bfseries 08}
  (2013) 090} [\href{https://arxiv.org/abs/1304.4926}{{\ttfamily 1304.4926}}].

\bibitem{Jacobson:1993xs}
T.~Jacobson and R.C.~Myers, \emph{{Black hole entropy and higher curvature
  interactions}},
  \href{https://doi.org/10.1103/PhysRevLett.70.3684}{\emph{Phys. Rev. Lett.}
  {\bfseries 70} (1993) 3684}
  [\href{https://arxiv.org/abs/hep-th/9305016}{{\ttfamily hep-th/9305016}}].

\bibitem{RyuTakayanagi1}
S.~Ryu and T.~Takayanagi, \emph{{Holographic derivation of entanglement entropy
  from AdS/CFT}},
  \href{https://doi.org/10.1103/PhysRevLett.96.181602}{\emph{Phys. Rev. Lett.}
  {\bfseries 96} (2006) 181602}
  [\href{https://arxiv.org/abs/hep-th/0603001}{{\ttfamily hep-th/0603001}}].

\bibitem{RyuTakayanagi2}
S.~Ryu and T.~Takayanagi, \emph{{Aspects of Holographic Entanglement Entropy}},
  \href{https://doi.org/10.1088/1126-6708/2006/08/045}{\emph{JHEP} {\bfseries
  08} (2006) 045} [\href{https://arxiv.org/abs/hep-th/0605073}{{\ttfamily
  hep-th/0605073}}].

\bibitem{DongExtremality}
X.~Dong and A.~Lewkowycz, \emph{{Entropy, Extremality, Euclidean Variations,
  and the Equations of Motion}},
  \href{https://doi.org/10.1007/JHEP01(2018)081}{\emph{JHEP} {\bfseries 01}
  (2018) 081} [\href{https://arxiv.org/abs/1705.08453}{{\ttfamily
  1705.08453}}].

\bibitem{Bhattacharyya2014}
A.~Bhattacharyya and M.~Sharma, \emph{{On entanglement entropy functionals in
  higher derivative gravity theories}},
  \href{https://doi.org/10.1007/JHEP10(2014)130}{\emph{JHEP} {\bfseries 10}
  (2014) 130} [\href{https://arxiv.org/abs/1405.3511}{{\ttfamily 1405.3511}}].

\bibitem{FursaevPatrushevSolodukhin}
D.V.~Fursaev, A.~Patrushev and S.N.~Solodukhin, \emph{{Distributional Geometry
  of Squashed Cones}},
  \href{https://doi.org/10.1103/PhysRevD.88.044054}{\emph{Phys. Rev. D}
  {\bfseries 88} (2013) 044054}
  [\href{https://arxiv.org/abs/1306.4000}{{\ttfamily 1306.4000}}].

\bibitem{Lovelock:1971yv}
D.~Lovelock, \emph{{The Einstein tensor and its generalizations}},
  \href{https://doi.org/10.1063/1.1665613}{\emph{J. Math. Phys.} {\bfseries 12}
  (1971) 498}.

\bibitem{BlackHolesHennigarMann}
R.A.~Hennigar and R.B.~Mann, \emph{{Black holes in Einsteinian cubic gravity}},
  \href{https://doi.org/10.1103/PhysRevD.95.064055}{\emph{Phys. Rev. D}
  {\bfseries 95} (2017) 064055}
  [\href{https://arxiv.org/abs/1610.06675}{{\ttfamily 1610.06675}}].

\bibitem{BlackHolesPablos}
P.~Bueno and P.A.~Cano, \emph{{Four-dimensional black holes in Einsteinian
  cubic gravity}},
  \href{https://doi.org/10.1103/PhysRevD.94.124051}{\emph{Phys. Rev. D}
  {\bfseries 94} (2016) 124051}
  [\href{https://arxiv.org/abs/1610.08019}{{\ttfamily 1610.08019}}].

\bibitem{AspectsHOG}
P.~Bueno, P.A.~Cano, V.S.~Min and M.R.~Visser, \emph{{Aspects of general
  higher-order gravities}},
  \href{https://doi.org/10.1103/PhysRevD.95.044010}{\emph{Phys. Rev. D}
  {\bfseries 95} (2017) 044010}
  [\href{https://arxiv.org/abs/1610.08519}{{\ttfamily 1610.08519}}].

\bibitem{Czech:2012bh}
B.~Czech, J.L.~Karczmarek, F.~Nogueira and M.~Van~Raamsdonk, \emph{{The Gravity
  Dual of a Density Matrix}},
  \href{https://doi.org/10.1088/0264-9381/29/15/155009}{\emph{Class. Quant.
  Grav.} {\bfseries 29} (2012) 155009}
  [\href{https://arxiv.org/abs/1204.1330}{{\ttfamily 1204.1330}}].

\bibitem{HubenySurfaces}
V.E.~Hubeny, \emph{{Extremal surfaces as bulk probes in AdS/CFT}},
  \href{https://doi.org/10.1007/JHEP07(2012)093}{\emph{JHEP} {\bfseries 07}
  (2012) 093} [\href{https://arxiv.org/abs/1203.1044}{{\ttfamily 1203.1044}}].

\bibitem{Mezei:2014zla}
M.~Mezei, \emph{{Entanglement entropy across a deformed sphere}},
  \href{https://doi.org/10.1103/PhysRevD.91.045038}{\emph{Phys. Rev. D}
  {\bfseries 91} (2015) 045038}
  [\href{https://arxiv.org/abs/1411.7011}{{\ttfamily 1411.7011}}].

\bibitem{Freedman:2016zud}
M.~Freedman and M.~Headrick, \emph{{Bit threads and holographic entanglement}},
  \href{https://doi.org/10.1007/s00220-016-2796-3}{\emph{Commun. Math. Phys.}
  {\bfseries 352} (2017) 407}
  [\href{https://arxiv.org/abs/1604.00354}{{\ttfamily 1604.00354}}].

\bibitem{Agon:2019qgh}
C.A.~Ag\'on and M.~Mezei, \emph{{Bit Threads and the Membrane Theory of
  Entanglement Dynamics}},  \href{https://arxiv.org/abs/1910.12909}{{\ttfamily
  1910.12909}}.

\bibitem{Agon:2020mvu}
C.A.~Ag\'on, E.~C\'aceres and J.F.~Pedraza, \emph{{Bit threads, Einstein's
  equations and bulk locality}},
  \href{https://arxiv.org/abs/2007.07907}{{\ttfamily 2007.07907}}.

\bibitem{Du:2019emy}
D.-H.~Du, C.-B.~Chen and F.-W.~Shu, \emph{{Bit threads and holographic
  entanglement of purification}},
  \href{https://doi.org/10.1007/JHEP08(2019)140}{\emph{JHEP} {\bfseries 08}
  (2019) 140} [\href{https://arxiv.org/abs/1904.06871}{{\ttfamily
  1904.06871}}].

\bibitem{Harper:2019lff}
J.~Harper and M.~Headrick, \emph{{Bit threads and holographic entanglement of
  purification}}, \href{https://doi.org/10.1007/JHEP08(2019)101}{\emph{JHEP}
  {\bfseries 08} (2019) 101}
  [\href{https://arxiv.org/abs/1906.05970}{{\ttfamily 1906.05970}}].

\bibitem{Bao:2019wcf}
N.~Bao, A.~Chatwin-Davies, J.~Pollack and G.N.~Remmen, \emph{{Towards a Bit
  Threads Derivation of Holographic Entanglement of Purification}},
  \href{https://doi.org/10.1007/JHEP07(2019)152}{\emph{JHEP} {\bfseries 07}
  (2019) 152} [\href{https://arxiv.org/abs/1905.04317}{{\ttfamily
  1905.04317}}].

\bibitem{Headrick:2017ucz}
M.~Headrick and V.E.~Hubeny, \emph{{Riemannian and Lorentzian flow-cut
  theorems}}, \href{https://doi.org/10.1088/1361-6382/aab83c}{\emph{Class.
  Quant. Grav.} {\bfseries 35} (2018) 10}
  [\href{https://arxiv.org/abs/1710.09516}{{\ttfamily 1710.09516}}].

\bibitem{Harper:2018sdd}
J.~Harper, M.~Headrick and A.~Rolph, \emph{{Bit Threads in Higher Curvature
  Gravity}}, \href{https://doi.org/10.1007/JHEP11(2018)168}{\emph{JHEP}
  {\bfseries 11} (2018) 168}
  [\href{https://arxiv.org/abs/1807.04294}{{\ttfamily 1807.04294}}].

\bibitem{Balasubramanian:2011ur}
V.~Balasubramanian, A.~Bernamonti, J.~de~Boer, N.~Copland, B.~Craps,
  E.~Keski-Vakkuri et~al., \emph{{Holographic Thermalization}},
  \href{https://doi.org/10.1103/PhysRevD.84.026010}{\emph{Phys. Rev. D}
  {\bfseries 84} (2011) 026010}
  [\href{https://arxiv.org/abs/1103.2683}{{\ttfamily 1103.2683}}].

\bibitem{Caceres:2012em}
E.~Caceres and A.~Kundu, \emph{{Holographic Thermalization with Chemical
  Potential}}, \href{https://doi.org/10.1007/JHEP09(2012)055}{\emph{JHEP}
  {\bfseries 09} (2012) 055} [\href{https://arxiv.org/abs/1205.2354}{{\ttfamily
  1205.2354}}].

\bibitem{Galante:2012pv}
D.~Galante and M.~Schvellinger, \emph{{Thermalization with a chemical potential
  from AdS spaces}}, \href{https://doi.org/10.1007/JHEP07(2012)096}{\emph{JHEP}
  {\bfseries 07} (2012) 096} [\href{https://arxiv.org/abs/1205.1548}{{\ttfamily
  1205.1548}}].

\bibitem{Zeng:2013mca}
X.~Zeng and W.~Liu, \emph{{Holographic thermalization in Gauss-Bonnet
  gravity}}, \href{https://doi.org/10.1016/j.physletb.2013.08.049}{\emph{Phys.
  Lett. B} {\bfseries 726} (2013) 481}
  [\href{https://arxiv.org/abs/1305.4841}{{\ttfamily 1305.4841}}].

\bibitem{Li:2013cja}
Y.-Z.~Li, S.-F.~Wu and G.-H.~Yang, \emph{{Gauss-Bonnet correction to
  Holographic thermalization: two-point functions, circular Wilson loops and
  entanglement entropy}},
  \href{https://doi.org/10.1103/PhysRevD.88.086006}{\emph{Phys. Rev. D}
  {\bfseries 88} (2013) 086006}
  [\href{https://arxiv.org/abs/1309.3764}{{\ttfamily 1309.3764}}].

\bibitem{Caceres:2015bkr}
E.~Caceres, M.~Sanchez and J.~Virrueta, \emph{{Holographic Entanglement Entropy
  in Time Dependent Gauss-Bonnet Gravity}},
  \href{https://doi.org/10.1007/JHEP09(2017)127}{\emph{JHEP} {\bfseries 09}
  (2017) 127} [\href{https://arxiv.org/abs/1512.05666}{{\ttfamily
  1512.05666}}].

\bibitem{Myers_2012}
R.C.~Myers and A.~Singh, \emph{{Entanglement Entropy for Singular Surfaces}},
  \href{https://doi.org/10.1007/JHEP09(2012)013}{\emph{JHEP} {\bfseries 09}
  (2012) 013} [\href{https://arxiv.org/abs/1206.5225}{{\ttfamily 1206.5225}}].

\bibitem{Bueno_2015}
P.~Bueno and R.C.~Myers, \emph{{Corner contributions to holographic
  entanglement entropy}},
  \href{https://doi.org/10.1007/JHEP08(2015)068}{\emph{JHEP} {\bfseries 08}
  (2015) 068} [\href{https://arxiv.org/abs/1505.07842}{{\ttfamily
  1505.07842}}].

\bibitem{Takahashi_2010}
T.~Takahashi and J.~Soda, \emph{{Master Equations for Gravitational
  Perturbations of Static Lovelock Black Holes in Higher Dimensions}},
  \href{https://doi.org/10.1143/PTP.124.911}{\emph{Prog. Theor. Phys.}
  {\bfseries 124} (2010) 911}
  [\href{https://arxiv.org/abs/1008.1385}{{\ttfamily 1008.1385}}].

\bibitem{BuenoVilar}
P.~Bueno, J.~Camps and A.~Vilar~L\'opez, \emph{{Holographic entanglement
  entropy for perturbative higher-curvature gravities}},
  \href{https://arxiv.org/abs/2012.14033}{{\ttfamily 2012.14033}}.

\end{thebibliography}\endgroup
\end{document}